\documentclass[12pt]{article}%

\begin{document}
\begin{center}{\large\bf A New Look at the Position Operator in Quantum Theory}\end{center}

\vskip 1em \begin{center} {\large Felix M. Lev} \end{center}
\vskip 1em \begin{center} {\it Artwork Conversion Software Inc.,
1201 Morningside Drive, Manhattan Beach, CA 90266, USA
(Email:  felixlev314@gmail.com)} \end{center}

\begin{flushleft}{\it Abstract:}\end{flushleft} 
The postulate that coordinate and momentum representations are related
to each other by the Fourier transform has been accepted from the beginning of quantum theory   
by analogy with classical electrodynamics.
As a consequence, an inevitable effect
in standard theory is the wave packet spreading (WPS) of the photon coordinate wave function in 
directions perpendicular to the photon momentum. 
This leads to several paradoxes. The most striking of them is that coordinate wave functions of 
photons emitted by stars have cosmic sizes and strong arguments indicate that this contradicts observational data.
We argue that the above postulate is based neither on strong 
theoretical arguments nor on experimental data and propose a new consistent definition
of the position operator. Then WPS in directions perpendicular to the particle
momentum is absent and the paradoxes are resolved. Different components of the new position 
operator do not commute with each other and,
as a consequence, there is no wave function in coordinate representation. Implications of the results
for entanglement, quantum locality and the problem of time in quantum theory are discussed.

\begin{flushleft} PACS: 11.30.Cp, 03.65.-w, 03.63.Sq, 03.65.Ta \end{flushleft}

\begin{flushleft} Keywords: quantum theory, position operator, semiclassical approximation\end{flushleft}

\section{Status of the position operator in quantum theory}
\label{intropos}
\subsection{Historical reasons for choosing standard form of position operator}
\label{historical}

It has been postulated from the beginning of quantum theory that the coordinate and momentum representations of wave functions are related to each other by the Fourier transform. One of the historical reasons was that in classical electrodynamics the coordinate and wave vector ${\bf k}$ representations are related analogously and we postulate that 
${\bf p}=\hbar {\bf k}$ where ${\bf p}$ is the particle momentum. Then, although the interpretations of classical fields on one hand and wave functions on the other are fully different, from mathematical point of view classical electrodynamics and quantum mechanics have much in common (and such a situation does not seem to be natural). 

Similarity of classical electrodynamics and quantum theory is reflected even in the terminology of the latter. 
The terms "wave function", "particle-wave duality" and "de Broglie wave length" have arisen at the beginning of 
quantum era in efforts to explain quantum behavior in terms of classical waves but now it is clear that no such 
explanation exists. The notion of wave
is purely classical; it has a physical meaning only as a way of describing systems of many particles by
their mean characteristics. In particular, such notions as frequency and wave length can be applied
only to classical waves, i.e. to systems consisting of many particles. If a particle state vector contains 
$exp[i({\bf p}{\bf r}-Et)/\hbar]$, where $E$ is the energy, then by analogy with the theory of
classical waves one might say that the particle is a wave with the frequency $\omega=E/\hbar$ and the 
(de Broglie) wave length $\lambda=2\pi\hbar/|{\bf p}|$. 
However, such defined quantities $\omega$ and $\lambda$ are not real frequencies
and wave lengths measured on macroscopic level. A striking example showing that on quantum level $\lambda$ does not have the usual meaning is that from the point of view of classical theory an electron having the size of the order of the
Bohr radius cannot emit a wave with $\lambda=21cm$ (this observation has been pointed out to me by Volodya Netchitailo).

In quantum theory the photon 
and other particles are characterized by their energies, momenta and 
other quantities for which there exist well defined operators while the notion of coordinates on quantum
level is a problem which is investigated in the present paper. The term "wave function" might be misleading
since in quantum theory it defines not amplitudes of waves but only amplitudes of probabilities. 
So, although in our opinion the term "state vector" is more pertinent than "wave function" we will use the
latter in accordance with the usual terminology, and the phrase that a photon has a frequency $\omega$ and
the wave length $\lambda$ will be understood only such that $\omega=E/\hbar$ and $\lambda=2\pi\hbar/|{\bf p}|$.

One of the examples of the above similarity follows. Consider a wave function of the form
$\psi({\bf r},t)=a({\bf r},t)exp[iS({\bf r},t)/\hbar]$, where $S({\bf r},t)$ is the classical action as a function of coordinates and time. Then
\begin{equation}
\frac{\partial \psi({\bf r},t)}{\partial {\bf r}}=[\frac{i}{\hbar}\frac{\partial 
S({\bf r},t)}{\partial {\bf r}}+\frac{1}{a({\bf r},t)}\frac{\partial 
a({\bf r},t)}{\partial {\bf r}}]\psi({\bf r},t)
\label{quasi}
\end{equation}
and analogously for $\partial \psi({\bf r},t)/\partial t$. In the formal limit
$\hbar\to 0$ the second term in the square brackets can be neglected and, as explained in textbooks on quantum mechanics (see e.g. Ref. \cite{LLIII})  the 
Schr\"{o}dinger equation becomes the Hamilton-Jacoby equation. This situation is 
analogous to the approximation of geometrical optics in classical electrodynamics 
(see e.g. Ref. \cite{LLII}) when fields contain a rapidly oscillating factor $exp[i\varphi ({\bf r},t)]$ where the function $\varphi({\bf r},t)$ is called eikonal. It satisfies the eikonal equation which coincides with the relativistic Hamilton-Jacobi equation for a particle with zero mass. This is reasonable
in view of the fact that electromagnetic waves consist of photons. 

Another example follows. In classical electrodynamics a wave packet moving even in empty space inevitably spreads out and this fact has been known
for a long time. For example, as pointed out by Schr\"{o}dinger (see pp. 41-44 in Ref. \cite{Schr}), in standard quantum mechanics a packet does not spread out if a particle is moving in a harmonic oscillator potential in contrast to 
"a wave packet in classical optics, which is dissipated in the course of time". 
However, as a consequence of the
similarity, a free quantum mechanical wave packet inevitably spreads out too. This effect is called wave packet spreading (WPS) and it is described in textbooks and many papers (see e.g. Refs. \cite{Dirac,QM} and references therein). In the present paper this effect is discussed in detail and we argue that
it plays a crucial role in drawing a conclusion on whether standard position operator is consistently
defined.

The requirement that the momentum and position operators are related to each other
by the Fourier transform is equivalent to standard commutation relations between these
operators and to the Heisenberg uncertainty principle (see Sec. \ref{classical}).

A reason for choosing standard form of the position operator is described, for example, in
the Dirac textbook \cite{Dirac}. Here Dirac argues that the momentum and position operators
should be such that their commutator should be proportional to the corresponding classical
Poisson bracket with the coefficient $i\hbar$. However, this argument is not convincing 
because only in very special
cases the commutator of two physical operators is a $c$-number. One can check, for example,
a case of momentum and position operators squared.

In Ref. \cite{Heisenberg} Heisenberg argues in favor of his principle by considering 
{\it Gedankenexperiment} with Heisenberg's microscope. Since that time the problem has been investigated 
in many publications. A discussion of the current status of the
problem can be found e.g. in Ref. \cite{Lahti} and references therein. A general opinion based on those 
investigations is   that Heisenberg's arguments are problematic but the uncertainty principle is valid, although several
authors argue whether standard mathematical notion of uncertainty (see Sec. \ref{classical})
is relevant for describing a real process of
measurement. However, a common assumption in those investigations
is that one can consider uncertainty relations for all the components of the position and momentum operators independently.
Below we argue that this assumption is not based on solid physical arguments. 

\subsection{Problem of consistency of standard position operator}
\label{consistency} 

Usual arguments in favor of choosing standard position and  momentum operators are that these 
operators have correct properties in semiclassical approximation (see e.g. Ref. \cite{LLIII}).
However, this requirement does not define the operator unambiguously. Indeed, if the operator
$B$ becomes zero in semiclassical limit then the operators $A$ and $A+B$ have the same
semiclassical limit. 

One of the principles of physics
is the correspondence one according to which any new theory should reproduce results of the
old well tested theory at some conditions. As noted above,
in the main approximation in $1/\hbar$ the Schr\"{o}dinger equation becomes the Hamilton-Jacoby equation if the coordinate wave function $\psi({\bf r},t)$ contains a factor $exp[iS({\bf r},t)/\hbar]$.
In textbooks this is usually treated as the correspondence principle between quantum 
and classical theories. However, the following question arises.

As follows from Eq. (\ref{quasi}),  the Hamilton-Jacoby equation is a good approximation for
the Schr\"{o}dinger equation if the index of the exponent changes much faster than the
amplitude $a({\bf r},t)$. Is this correct to define semiclassical approximation by this condition?
Quantum theory fully reproduces the results of classical one when not only this condition is
satisfied but, in addition, the amplitude has a sharp maximum along the classical
trajectory. If the latter is true at some moment of time then, in view of the WPS effect, one
cannot gurantee that this will be true always.

At the beginning of quantum theory the WPS effect has been investigated by de Broglie,
Darwin and Schr\"{o}dinger. The fact that WPS is inevitable has been treated by several authors as unacceptable
and as an indication that standard quantum theory should be modified. For example, de Broglie has proposed to
describe a free particle not by the Schr\"{o}dinger equation but by a wavelet which satisfies a nonlinear equation
and does not spread out (a detailed description of de Broglie's wavelets can be found e.g. in Ref. \cite{Barut}).
Sapogin writes (see Ref. \cite{Sapogin} and references therein) that "Darwin showed 
that such packet quickly and steadily dissipates
and disappears" and proposes an alternative to standard theory which he calls unitary unified
quantum field theory. 

At the same time, it has not been explicitly shown that numerical results on WPS are 
incompatible with experimental data. For example, it is known (see Sec. \ref{NRWPS})
that for macroscopic bodies the effect of WPS is extremely small. Probably it is also believed that in 
experiments on the Earth with atoms 
and elementary particles spreading does not have enough time to manifest itself although we have not 
found an explicit statement on this problem in the literature. 
According to our observations, different physicists have different opinions on the role of
WPS in different phenomena but in any case the absolute majority of physicists do not treat WPS as a drawback of the theory. 

A natural problem arises what happens to photons which can travel from distant objects to Earth even for 
billions of years. As shown in Sec. \ref{experiment}, standard theory predicts that, as a consequence
of WPS, wave functions of such photons will have the size of the order of millions or billions
kilometers or even more. Does this contradict observations? We argue that it does and the reason of 
the paradox is that 
standard position operator is not consistently defined. Hence the inconsistent definition of the position 
operator is not only an academic problem but leads to the above paradox.
 
In view of the fact that the coordinate and momentum representations are related to each other by the
Fourier transform, one might think that the position and momentum operators are on equal footing. However,
this is not the case for the following reasons. In quantum theory each
elementary particle is described by an irreducible representation (IR) of the symmetry algebra. 
For example, in Poincare invariant theory the set of momentum operators represents three of ten 
linearly independent representation
operators of the Poincare algebra and hence those operators are consistently defined.
On the other hand, among the representation operators there is no position operator. 
So the assumption that the position operator in momentum representation is $i\hbar\partial /\partial {\bf p}$ should be substantiated.

Consider first a one-dimensional case. As argued in textbooks (see e.g. Ref. \cite{LLIII}),
if the mean value of the $x$ component of the momentum $p_x$ is rather large, the definition of the coordinate operator 
$i\hbar \partial/\partial p_x$ can be justified  but this definition does not have a physical meaning in situations when $p_x$ is small. 
This is clear even from the fact that if $p_x$ is small then $exp(ip_xx/\hbar)$ is not a rapidly oscillating function of $x$.

Consider now the three-dimensional case.
If all the components $p_j$ ($j=1,2,3$) are rather large then all the operators 
$i\hbar \partial/\partial p_j$ can have a physical meaning. A semiclassical wave function $\chi({\bf p})$ in
momentum space should describe a narrow distribution around the mean value ${\bf p}_0$. Suppose now that coordinate 
axes are chosen such ${\bf p}_0$ is directed along the $z$ axis. Then the mean values of the $x$ and $y$
components of the momentum operator equal zero and
the operators $i\hbar \partial/\partial p_j$ cannot be physical for $j=1,2$, i.e. in directions perpendicular
to the particle momentum. 
The situation when a definition of an operator is physical or not depending on the choice of coordinate axes is not acceptable. Hence standard definition of the position operator is not physical.

\subsection{When do we need position operator in quantum theory?}
\label{when}

The position operator is used in many standard problems of quantum theory. For example,
one of the arguments in favor of its validity  is that the nonrelativistic 
Schr\"{o}dinger equation correctly describes the hydrogen energy levels, the Dirac equation correctly describes fine structure corrections to these levels etc. Historically these equations have been first written in coordinate space and in textbooks they are still discussed in this form. However, from the point of view of the present knowledge those equations should be treated as follows.

A fundamental theory describing electromagnetic interactions on quantum level is quantum electrodynamics (QED). This theory proceeds from quantizing classical Lagrangian which is only an auxiliary tool for constructing S-matrix. 
The argument ${\bf x}$ in the Lagrangian density $L(t, {\bf x})$ cannot be treated as a position operator because
$L(t, {\bf x})$ is constructed from field functions which do not have a probabilistic interpretation. When
quantization is accomplished, the results of QED are formulated exclusively in momentum space and the theory does not contain space-time at all. 

In particular, as follows from Feynman diagrams for the one-photon exchange, in the approximation $(v/c)^2$  the electron in the hydrogen atom can be described in the potential formalism where the potential acts on the wave function in momentum space.  So for calculating energy levels one should solve the eigenvalue problem for the Hamiltonian with this potential. This is an integral equation which can be solved by different methods. One of the convenient methods is to apply the Fourier transform and get standard Schr\"{o}dinger or Dirac equation in coordinate representation with the Coulomb potential. Hence the fact that the results for energy levels are in good agreement with experiment shows only that QED defines the potential correctly and {\it standard coordinate Schr\"{o}dinger and Dirac equations are only convenient mathematical ways of solving the eigenvalue problem}. For this problem the physical meaning of the position operator is not important at all. One can consider other transformations of the original integral equation and define other position operators. The fact that for non-standard choices one might obtain something different from the Coulomb potential is not important on quantum level. On classical level the interaction between two charges can be described by the Coulomb potential but this does not imply that on quantum level the potential in coordinate representation should be necessarily Coulomb.

Let us also note the following. In the literature the statement that the Coulomb law works with a high accuracy is often 
substantiated from the point of view that predictions of QED have been experimentally confirmed with a high accuracy. 
However, as follows from the above remarks, the meaning of distance on quantum level is not
clear and in QED the law $1/r^2$ can be tested only if we
assume additionally that the coordinate and momentum representations are related to each other by the Fourier transform. 
So a conclusion about the validity of the law can be made only on the basis of macroscopic experiments.
A conclusion made from the results of classical Cavendish and Maxwell experiments is that if the exponent in Coulomb's
law is not 2 but $2\pm q$ then $q<1/21600$. The accuracy of those experiments have been considerably improved in
the experiment \cite{Plimpton} the result of which is $q<2\cdot 10^{-9}$. However, the
Cavendish-Maxwell experiments and the experiment \cite{Plimpton} do not involve pointlike electric charges. 
Cavendish and Maxwell used a 
spherical air condenser consisting of two insulated spherical shells while the authors of Ref. \cite{Plimpton}
developed a technique where the difficulties due to spontaneous ionization and contact potentials were avoided.
Therefore the conclusion that $q<2\cdot 10^{-9}$ for pointlike electric charges requires additional assumptions.  

Another example follows. It is said that the spatial distribution of the electric charge inside a system
can be extracted from measurements of form-factors in the electron scattering on this system. 
However, the information about the experiment is again given only in terms of
momenta and conclusions about the spatial distribution can be drawn only if we assume additionally how
the position operator is expressed in terms of momentum variables. On quantum level the physical meaning of
such a spatial distribution is not fundamental.

In view of the above discussion, since the {\it results} of existing fundamental quantum theories 
describing interactions on quantum level (QED, electroweak theory and QCD) are formulated exclusively
in terms of the S-matrix in momentum space without any mentioning of space-time, {\it for investigating such
stationary quantum problems as calculating energy levels, form-factors etc., the notion of the
position operator is not needed}.

However, the choice of the position operator is important in nonstationary problems when evolution is
described by the time dependent Schr\"{o}dinger equation (with the nonrelativistic or relativistic Hamiltonian). As follows 
from the correspondence principle, 
 quantum theory should reproduce the motion of a particle along
the classical trajectory defined by classical equations of motion. Hence the position operator is needed only
in semiclassical approximation and it should be {\it defined} from additional considerations. 

In standard approaches to quantum theory the existence of space-time background is assumed from the beginning.
Then the position operator for a particle in this background is the operator of multiplication by the
particle radius-vector ${\bf r}$.
As explained in textbooks on quantum mechanics (see e.g. Ref. \cite{LLIII}), the result 
$-i\hbar \partial/\partial{\bf r}$
for the momentum operator can be justified from the requirement that quantum theory should correctly reproduce classical
results in semiclassical approximation. However, as noted above, this requirement does not define the operator 
unambiguously.

A standard approach to Poincare symmetry on quantum level follows.  
Since Poincare group is the group of motions of Minkowski space, quantum states should be described by representations of 
the Poincare group. In turn, this implies that the representation generators should commute according to the commutation 
relations of the Poincare group Lie algebra:
\begin{eqnarray}
&&[P^{\mu},P^{\nu}]=0\quad [P^{\mu},M^{\nu\rho}]=-i(\eta^{\mu\rho}P^{\nu}-\eta^{\mu\nu}P^{\rho})\nonumber\\
&&[M^{\mu\nu},M^{\rho\sigma}]=-i (\eta^{\mu\rho}M^{\nu\sigma}+\eta^{\nu\sigma}M^{\mu\rho}-
\eta^{\mu\sigma}M^{\nu\rho}-\eta^{\nu\rho}M^{\mu\sigma})
\label{PCR}
\end{eqnarray}
where $P^{\mu}$ are the operators of the four-momentum, $M^{\mu\nu}$ are the operators of
Lorentz angular momenta, the diagonal metric tensor $\eta^{\mu\nu}$ has the nonzero components
$\eta^{00}=-\eta^{11}=-\eta^{22}=-\eta^{33}=1$ and $\mu,\nu=0,1,2,3$. It is usually said that the above
relations are written in the system of units $c=\hbar=1$. However, as we argue in Ref. \cite{symm1401}, quantum
theory should not contain $c$ and $\hbar$ at all; those quantities arise only because we wish to measure
velocities in $m/s$ and angular momenta in $kg\times m^2/s$.

The above approach is in the spirit of Klein's Erlangen program in mathematics.
However, as we argue in Refs. \cite{symm1401,DS}, quantum theory should not be based on 
classical space-time background. The notion of space-time background contradicts the basic principle of
physics that a definition of a physical quantity is a description of how this quantity should be measured.
Indeed one cannot measure coordinates of a manifold which exists only in our imagination.

As we argue in Refs. \cite{symm1401,DS} and other publications, the approach should be the opposite. 
Each system is described by a set of independent operators.
By definition, the rules how these operators commute with each other define the symmetry algebra. 
In particular, {\it by definition}, Poincare symmetry on quantum level means that the operators commute
according to Eq. (\ref{PCR}). This definition does not involve Minkowski space at all.
Such a definition of symmetry on quantum level is in the spirit of Dirac's paper \cite{Dir}.

The fact that an elementary particle in quantum theory is described by an IR of the symmetry algebra can
be treated as a definition of the elementary particle. In Poincare invariant theory the IRs can be
implemented in a space of functions $\chi({\bf p})$ such that $\int |\chi({\bf p})|^2d^3{\bf p}<\infty$ 
(see Sec. \ref{momentum}).
In this representation the momentum operator ${\bf P}$ is defined {\it unambiguously} and is simply the operator of multiplication by ${\bf p}$.
A standard {\it assumption} is that the position operator in this representation is $i\hbar \partial/\partial {\bf p}$. However, as argued above, this assumption is not consistent.

In the present paper we propose a consistent 
definition of the position operator. As a consequence, in our approach WPS in directions perpendicular to the particle
momentum is absent regardless of whether the particle is nonrelativistic or relativistic. 
Moreover, for an ultrarelativistic particle the effect of 
WPS is absent at all. In our approach different components of the position operator do not commute with each other and,
as a consequence, there is no wave function in coordinate representation. 

Our presentation is self-contained and for reproducing the results of the calculations no special knowledge
is needed. The paper is organized as follows. In Secs. \ref{classical} and \ref{momentum} we discuss the 
approach to the position operator in standard nonrelativistic and relativistic quantum theory, respectively. 
An inevitable consequence 
of this approach is the effect of WPS of the coordinate
wave function which is discussed in Secs. \ref{NRWPS} and \ref{RelWPS} for the nonrelativistic and
relativistic cases, respectively. In Sec. \ref{WPW} we discuss a relation between the WPS effects for a classical
wave packet and for photons comprising this packet. In Sec. \ref{coherent} the problem of WPS in coherent states is
discussed. In Sec. \ref{experiment} we show that the WPS effect leads to several paradoxes and,  
as discussed in Sec. \ref{Discussion}, in standard theory it is not possible to avoid those paradoxes.
Our approach to a consistent definition of the position operator and its application to WPS are discussed 
in Secs. \ref{consistent}-\ref{newWPS}. Finally, in Sec. \ref{conclusion} we discuss implications of the results
for entanglement, quantum locality and the problem of time in quantum theory.

\begin{sloppypar}
\section{Position operator in nonrelativistic quantum mechanics}
\label{classical}
\end{sloppypar}

In quantum theory, states of a system are represented by elements of a projective Hilbert space. The fact that a
Hilbert space $H$ is projective means that if $\psi\in H$ is a state then $const\cdot\psi$ is the same state.
The matter is that not the probability itself but only relative probabilities of different measurement outcomes 
have a physical meaning. In this paper we will work with states $\psi$ normalized to one, i.e. such that 
$||\psi||=1$ where $||...||$ is a norm. It is defined such that if 
 $(...,...)$ is a scalar product in $H$ then $||\psi||=(\psi,\psi)^{1/2}$.  

In quantum theory every physical quantity is described by a selfadjoint operator. 
Each selfadjoint operator is
Hermitian i.e. satisfies the property $(\psi_2,A\psi_1)=(A\psi_2,\psi_1)$ for any states belonging to the
domain of $A$. If $A$ is an operator of some
quantity then the mean value of the quantity and its uncertainty in state $\psi$ are given by ${\bar A}=(\psi,A\psi)$
and $\Delta A=||(A-{\bar A})\psi||$, respectively. The condition that a quantity corresponding to the operator $A$
is semiclassical in state $\psi$ can be defined such that $\Delta A\ll |{\bar A}|$. This implies that the
quantity can be semiclassical only if $|{\bar A}|$ is rather large. In particular, if ${\bar A}=0$ then the
quantity cannot be semiclassical.

Let $B$ be an operator corresponding to another physical quantity and ${\bar B}$ and $\Delta B$ be the
mean value and the uncertainty of this quantity, respectively. We can write $AB=\{A,B\}/2 + [A,B]/2$
where the commutator $[A,B]=AB-BA$ is anti-Hermitian and the anticommutator $\{A,B\}=AB+BA$ is Hermitian. 
Let $[A,B]=-iC$ and ${\bar C}$ be the mean value of the operator $C$.

A question arises whether two physical
quantities corresponding to the operators $A$ and $B$ can be simultaneously semiclassical in state $\psi$. Since
$||\psi_1||||\psi_2||\geq |(\psi_1,\psi_2)|$, we have that
\begin{equation}
\Delta A \Delta B\geq \frac{1}{2}|(\psi, (\{A-{\bar A},B-{\bar B}\}+[A,B])\psi)|
\end{equation}
Since $(\psi,\{A-{\bar A},B-{\bar B}\}\psi)$ is real and $(\psi,[A,B]\psi)$ is imaginary, we get
\begin{equation}
\Delta A \Delta B \geq \frac{1}{2}|{\bar C}|
\label{uncert}
\end{equation}
This condition is known as a general uncertainty relation between two quantities. A well-known special case is
that if $P$ is the $x$ component of the momentum operator and $X$ is the operator of multiplication by $x$
then $[P,X]=-i\hbar$ and $\Delta p \Delta x\geq \hbar/2$.
The states where $\Delta p \Delta x= \hbar/2$ are called coherent ones. They are treated such that the momentum
and the coordinate are simultaneously semiclassical in a maximal possible extent. A well-known example is that if
$$\psi(x)=\frac{1}{a^{1/2}\pi^{1/4}}exp[\frac{i}{\hbar}p_0x-\frac{1}{2a^2}(x-x_0)^2]$$
then ${\bar X}=x_0$, ${\bar P}=p_0$, $\Delta x=a/\sqrt{2}$ and $\Delta p=\hbar /(a\sqrt{2})$.

Consider first a one dimensional motion. In standard textbooks on quantum mechanics, the presentation
starts with a wave function $\psi(x)$ in coordinate space since it is implicitly assumed that the meaning of
space coordinates is known. Then a question arises why $P=-i\hbar d/dx$ should be treated as the momentum operator.
The explanation follows.

Consider wave functions having the form $\psi(x)=exp(ip_0x/\hbar)a(x)$ where the amplitude $a(x)$ has a sharp maximum 
near $x=x_0\in [x_1,x_2]$ such that $a(x)$ is not small only when $x\in [x_1,x_2]$. Then $\Delta x$ is of the
order $x_2-x_1$ and the condition that the coordinate is semiclassical is $\Delta x\ll |x_0|$. 
Since $-i\hbar d\psi(x)/dx=p_0\psi(x)-i\hbar exp(ip_0x/\hbar)da(x)/dx$, we see that $\psi(x)$ will be approximately
the eigenfunction of $-i\hbar d/dx$ with the eigenvalue $p_0$ if $|p_0a(x)|\gg \hbar|da(x)/dx|$. Since 
$|da(x)/dx|$ is of the order of $|a(x)/\Delta x|$, we have a condition
$|p_0\Delta x| \gg \hbar$. Therefore if the momentum operator is $-i\hbar d/dx$, the uncertainty of momentum $\Delta p$
is of the order of $\hbar/\Delta x$, $|p_0|\gg \Delta p$ and this implies that the momentum is also semiclassical. 
At the same time, $|p_0\Delta x|/2\pi\hbar$ is approximately the number of oscillations which the exponent 
makes on the segment $[x_1,x_2]$. Therefore the number of oscillations should be much greater than unity. In particular, 
semiclassical approximation cannot be valid if $\Delta x$ is very small, but on the other hand, $\Delta x$ cannot be
very large since it should be much less than $x_0$. Another justification of the fact that $-i\hbar d/dx$ is the
momentum operator is that in the formal limit $\hbar\to 0$ the Schr\"{o}dinger equation becomes the Hamilton-Jacobi
equation. 

We conclude that the choice of $-i\hbar d/dx$ as the momentum operator is justified from the requirement that
in semiclassical approximation this operator becomes the classical momentum. However, it is obvious that this
requirement does not define the operator uniquely: any operator ${\tilde P}$ such that ${\tilde P}-P$ disappears
in semiclassical limit, also can be called the momentum operator.

One might say that the choice $P=-i\hbar d/dx$ can also be justified from the following considerations. In nonrelativistic
quantum mechanics we assume that the theory should be invariant under the action of the Galilei group, which is a group
of transformations of Galilei space-time. The $x$ component of the momentum operator should be the generator corresponding
to spatial translations along the $x$ axis and $-i\hbar d/dx$ is precisely the required operator. In this consideration
one assumes that the space-time background has a physical meaning while, as discussed in Refs. \cite{symm1401,DS} and references therein, this is not the case.

As noted in Refs. \cite{symm1401,DS} and references therein, one should start not from space-time but from a 
symmetry algebra. Therefore in nonrelativistic
quantum mechanics we should start from the Galilei algebra and consider its IRs. 
For simplicity we again consider a
one dimensional case. Let $P_x=P$ be one of representation operators in an IR of the Galilei algebra. We can implement
this IR in a Hilbert space of functions $\chi(p)$ such that 
$\int_{-\infty}^{\infty}|\chi(p)|^2dp < \infty$ and $P$ is
the operator of multiplication by $p$, i.e. $P\chi(p)=p\chi(p)$. 
Then a question arises how the operator of the $x$
coordinate should be defined. In contrast to the momentum operator, the 
coordinate one is not defined by the representation and so it should be defined 
from additional assumptions. Probably a future quantum theory of
measurements will make it possible to construct operators of physical quantities 
from the rules how these quantities should be measured. However, at present 
we can construct necessary operators only from rather intuitive considerations. 

By analogy with the above discussion, one can say that semiclassical wave functions should
be of the form $\chi(p)=exp(-ix_0p/\hbar)a(p)$ where the amplitude $a(p)$ has a sharp maximum 
near $p=p_0\in [p_1,p_2]$ such that $a(p)$ is not small only when $p\in [p_1,p_2]$. Then $\Delta p$ is of the
order of $p_2-p_1$ and the condition that the momentum is semiclassical is $\Delta p\ll |p_0|$. 
Since $i\hbar d\chi(p)/dp=x_0\chi(p)+i\hbar exp(-ix_0p/\hbar)da(p)/dp$, we see that $\chi(p)$ will be approximately
the eigenfunction of $i\hbar d/dp$ with the eigenvalue $x_0$ if $|x_0a(p)|\gg \hbar|da(p)/dp|$. Since 
$|da(p)/dp|$ is of the order of $|a(p)/\Delta p|$, we have a condition
$|x_0\Delta p| \gg \hbar$. Therefore if the coordinate operator is $X=i\hbar d/dp$, the uncertainty of coordinate $\Delta x$
is of the order of $\hbar/\Delta p$, $|x_0|\gg \Delta x$ and this implies that the coordinate defined in such a way
is also semiclassical. We can also note that $|x_0\Delta p|/2\pi\hbar$ is approximately the number of oscillations which the exponent 
makes on the segment $[p_1,p_2]$ and therefore the number of oscillations should be much greater than unity. 
It is also clear that semiclassical approximation cannot be valid if $\Delta p$ is very small, but on the other 
hand, $\Delta p$ cannot be very large since it should be much less than $p_0$. 
By analogy with the above discussion,
the requirement that the operator $i\hbar d/dp$ becomes the coordinate in classical limit does not define the
operator uniquely. In nonrelativistic quantum mechanics it is assumed that the coordinate is a well defined
physical quantity even on quantum level and that $i\hbar d/dp$ is the most pertinent choice. 

The above results can be formally generalized to the three-dimensional case. For example, if the coordinate wave
function is chosen in the form
\begin{equation}
\psi({\bf r})=\frac{1}{\pi^{3/4}a^{3/2}}exp[-\frac{({\bf r}-{\bf r}_0)^2}{2a^2}+\frac{i}{\hbar}{\bf p}_0{\bf r}]
\label{psir}
\end{equation}
then the momentum wave function is
\begin{equation}
\chi({\bf p})=\int exp(-\frac{i}{\hbar}{\bf p}{\bf r})\psi({\bf r})\frac{d^3{\bf r}}{(2\pi\hbar)^{3/2}}=
\frac{a^{3/2}}{\pi^{3/4}\hbar^{3/2}}exp[-\frac{({\bf p}-{\bf p}_0)^2a^2}{2\hbar^2}-
\frac{i}{\hbar}({\bf p}-{\bf p}_0){\bf r}_0]
\label{chip}
\end{equation}
It is easy to verify that
\begin{equation}
||\psi||^2=\int |\psi({\bf r})|^2d^3{\bf r}=1,\quad ||\chi||^2=\int |\chi({\bf p})|^2d^3{\bf p}=1,
\label{norm}
\end{equation}
the uncertainty of each component of the coordinate operator is $a/\sqrt{2}$ and the uncertainty of each
component of the momentum operator is $\hbar /(a\sqrt{2})$. Hence one might think that Eqs. (\ref{psir}) 
and (\ref{chip}) describe a state which is semiclassical in a maximal possible extent.

Let us make the following remark about semiclassical vector quantities. We defined a quantity as semiclassical if
its uncertainty is much less than its mean value. In particular, as noted above, a quantity cannot be
semiclassical if its mean value is small. In the case of vector quantities we have sets of three physical quantities.
Some of them can be small and for them it is meaningless to discuss whether they are semiclassical or not. We say that
a vector quantity is semiclassical if all its components which are not small are semiclassical and there should be
at least one semiclassical component. 

For example, if the mean value of the momentum ${\bf p}_0$ is directed along the $z$ axes then the $xy$ components of the
momentum are not semiclassical but the three-dimensional vector quantity ${\bf p}$ can be semiclassical
if ${\bf p}_0$ is rather large. However, in that case the definitions of the $x$ and $y$ components of the position
operator as $x=i\hbar \partial/\partial p_x$ and $y=i\hbar \partial/\partial p_y$ become inconsistent.
The situation when the validity of an operator depends on the choice of directions of the coordinate axes
is not acceptable and hence the above definition of the position operator is at least problematic.  

Let us note that semiclassical states can be constructed not only in momentum or coordinate representations.
For example, instead of momentum wave functions $\chi({\bf p})$ one can work in the representation where 
the quantum numbers $(p,l,\mu)$ in wave functions $\chi(p,l,\mu)$ mean the magnitude of the momentum $p$, the
orbital quantum number $l$ (such that a state is the eigenstate of the orbital momentum squared ${\bf L}^2$
with the eigenvalue $l(l+1)$) and the magnetic quantum number $\mu$ (such that a state is the eigenvector or
$L_z$ with the eigenvalue $\mu$). A state described by a $\chi(p,l,\mu)$ will be semiclassical with respect to those quantum
numbers if $\chi(p,l,\mu)$ has a sharp maximum at $p=p_0$, $l=l_0$, $\mu=\mu_0$ and the widths of the maxima in
$p$, $l$ and $\mu$ are much less than $p_0$, $l_0$ and $\mu_0$, respectively. However, by analogy with the above
discussion, those widths cannot be arbitrarily small if one wishes to have other semiclassical variables
(e.g. the coordinates). Examples of such situations will be discussed in Sec. \ref{newsemicl}.

\section{Wave packet spreading in nonrelativistic quantum mechanics}
\label{NRWPS}

As noted by Pauli (see p. 63 of Ref. \cite{Pauli}), at early stages of quantum theory some
authors treated time $t$ as the operator commuting with the Hamiltonian as $[H,t] = i\hbar$
but such a treatment is not correct. For example, one cannot construct the eigenstate
of the time operator with the eigenvalue 5000 BC or 3000 AD. Hence the quantity
$t$ can be only a classical parameter (see also Ref. \cite{time}). We see that the principle of
quantum theory that every physical quantity is defined by an operator does not apply
to time. The problem of time in quantum theory is discussed in a wide literature and 
remarks on this problem are made in Sec. \ref{conclusion}. However, for now we assume
that standard treatment of time is valid, i.e. that time is a classical parameter such
that the dependence of the wave function on time is defined by the Hamiltonian
according to the Schr\"{o}dinger equation.

In nonrelativistic quantum mechanics the Hamiltonian of a free particle with the mass $m$ is $H={\bf p}^2/2m$ and hence, 
as follows from Eq. (\ref{chip}), in the model discussed above the dependence of the momentum wave function on $t$ is 
\begin{equation}
\chi({\bf p}, t)=\frac{a^{3/2}}{\pi^{3/4}\hbar^{3/2}}exp[-\frac{({\bf p}-{\bf p}_0)^2a^2}{2\hbar^2}-
\frac{i}{\hbar}({\bf p}-{\bf p}_0){\bf r}_0-\frac{i{\bf p}^2t}{2m\hbar}]
\label{chipt}
\end{equation}
It is easy to verify that for this state the mean value of the operator ${\bf p}$ and the uncertainty of each 
momentum component are the same as for the state $\chi({\bf p})$, i.e. those quantities do not change with time.

Consider now the dependence of the coordinate wave function on $t$. This dependence can be calculated by
using Eq. (\ref{chipt}) and the fact that
\begin{equation}
\psi({\bf r},t)=\int exp(\frac{i}{\hbar}{\bf p}{\bf r})\chi({\bf p},t)\frac{d^3{\bf p}}{(2\pi\hbar)^{3/2}}
\label{Fourier}
\end{equation}
The result of a direct calculation is
\begin{equation}
\psi({\bf r},t)=\frac{1}{\pi^{3/4}a^{3/2}}(1+\frac{i\hbar t}{ma^2})^{-3/2}exp[-\frac{({\bf r}-{\bf r}_0-{\bf v}_0t)^2}{2a^2(1+\frac{\hbar^2t^2}{m^2a^4})}(1-\frac{i\hbar t}{ma^2})+\frac{i}{\hbar}{\bf p}_0{\bf r}-\frac{i{\bf p}_0^2t}{2m\hbar}]
\label{psirt}
\end{equation}
where ${\bf v}_0={\bf p}_0/m$ is the classical velocity. This result shows that the semiclassical wave packet is
moving along the classical trajectory ${\bf r}(t)={\bf r}_0+{\bf v}_0t$. At the same time, it is now obvious that
the uncertainty of each coordinate depends on time as 
\begin{equation}
\Delta x_j(t)=\Delta x_j(0)(1+\hbar^2t^2/m^2a^4)^{1/2}, \quad (j=1,2,3)
\label{deltax}
\end{equation}
where $\Delta x_j(0)=a/\sqrt{2}$, i.e. the width of the
wave packet in coordinate representation is increasing. This fact, known as the wave-packet spreading (WPS), is 
described in many textbooks and papers (see e.g. the textbooks \cite{Dirac,QM} and references therein).
It shows that
if a state was semiclassical in the maximal extent at $t=0$, it will not have this property at $t>0$ and the
accuracy of semiclassical approximation will decrease with the increase of $t$. The characteristic time of spreading
can be defined as $t_*=ma^2/\hbar$. For macroscopic bodies this is an extremely large quantity
and hence in macroscopic physics the WPS effect can be neglected. In the formal limit
$\hbar\to 0$, $t_*$ becomes infinite, i.e. spreading does not take place. This shows that WPS is a pure
quantum phenomenon. For the first time the result (\ref{psirt}) has been obtained by Darwin in Ref. \cite{Darwin}.

One might pose a problem whether the WPS effect is specific only for Gaussian wave functions. One might
expect that this effect will take place in general situations since each component of the standard position
operator $i\hbar \partial/\partial {\bf p}$ does not commute with the Hamiltonian and so the distribution of the 
corresponding physical quantity will be time dependent. A good example showing inevitability of WPS follows.
If at $t=0$ the coordinate wave function is $\psi_0({\bf r})$ then, as follows from Eqs. (\ref{chip}) and (\ref{Fourier}),
\begin{equation}
\psi({\bf r},t)=\int exp\{\frac{i}{\hbar}[{\bf p}({\bf r}-{\bf r}')-\frac{{\bf p}^2t}{2m}]\}\psi_0({\bf r}')
\frac{d^3{\bf r}'d^3{\bf p}}{(2\pi\hbar)^3}
\label{tnonrel}
\end{equation}
As follows from this expression, if $\psi_0({\bf r})\neq 0$ only if ${\bf r}$ belongs to a finite vicinity of some 
vector ${\bf r}_0$ then at any $t>0$ the support of $\psi({\bf r},t)$ belongs to the whole three-dimensional space, i.e. the
wave function spreads out with an infinite speed. One might think that in nonrelativistic theory this is not
unacceptable since this theory can be treated as a formal limit $c\to\infty$ of relativistic theory. In the next sections 
we will discuss an analogous situation in relativistic theory.

As shown in Ref. \cite{Berry} titled "Nonspreading wave packets", for a one-dimensional wave function in the form of an Airy function, spreading does not take place and the maximum of the quantity $|\psi(x)|^2$ propagates with constant acceleration even in the absence of external forces. Those properties of Airy packets have been observed in optical experiments \cite{Siviloglou}. However, since such a wave function is not normalizable, the term "wave packet" in the given situation might be misleading since the mean values and uncertainties of the coordinate and momentum cannot be calculated in a standard way. Such a wave function can be constructed only in a limited region of space. As explained in Ref. \cite{Berry}, this wave function describes not a particle but rather families of particle orbits. As shown in Ref. \cite{Berry}, one can construct a normalized state which is a superposition of Airy functions with Gaussian coefficients and "eventually the spreading due to the Gaussian cutoff takes over". This is an additional argument that the effect of WPS is an
inevitable consequence of standard quantum theory.

Since quantum theory is invariant under time reversal, one might ask the following question: is it possible that the width of the wave packet in coordinate representation is decreasing with time? From the formal point of view,
the answer is "yes". Indeed, the solution given by Eq. (\ref{psirt}) is valid not only when $t\geq 0$ but when
$t < 0$ as well. Then, as follows from Eq. (\ref{deltax}), the uncertainty of each coordinate is decreasing when $t$
changes from some negative value to zero. However, eventually the value of $t$ will become positive and the quantities
$\Delta x_j(t)$ will grow to infinity. In the present paper we consider situations when a photon is created on
atomic level and hence one might expect that its initial coordinate uncertainties are not large. However, when the
photon travels a long distance to Earth, those uncertainties become much greater, i.e. the term WPS reflects the
physics adequately. 

\section{Mott-Heisenberg problem and its generalization}
\label{Mott}

In 1929 Mott and Heisenberg considered the following problem. Let an alpha-particle 
be emitted by a nucleus in a radioactive decay. Suppose, for simplicity, that the particle
has been emitted in a state with zero angular momentum. Then the momentum wave function is spherically symmetric and all directions of the momentum have equal probabilities. However, when the
particle is detected in Wilson's cloud chamber, the registered trajectory is always linear  as if
the particle moved along a classical trajectory. The explanation of the paradox has been 
given in Ref. \cite{Mott}. In this section we consider a general case when it is not assumed 
that the partical wave function is spherically symmetric. 


Consider the state (\ref{tnonrel}) after a long period of time such that $D\gg a$ where
$D=\hbar t/(ma)$. As follows from Eq. (\ref{tnonrel}), at this condition the width of the
coordinate wave function is of the order $D$. Suppose that the particle is emitted at the
origin such that ${\bf r}_0=0$. Suppose that a measuring device is at the point ${\bf r}_1$
and the size of the device is of the order of $d$. Although the device is macroscopic, we assume
that $D$ is already so large that $D\gg d$. A problem arises at which momentum range the
particle will be detected.

For solving this problem we first project the coordinate wave function onto the region of space
belonging to the device. Assume that the projected wave function is
\begin{equation}
{\tilde\psi({\bf r},t)}=exp[-\frac{({\bf r}-{\bf r}_1)^2}{2d^2}]\psi({\bf r},t)
\label{project}
\end{equation}
A direct calculation shows that the norm of this state is
\begin{equation}
||{\tilde\psi}||^2=(\frac{d}{D})^3exp[-\frac{({\bf r}_1-{\bf v}_0t)^2}{D^2}]
\label{projnorm}
\end{equation}
This result is obvious because the wave function of the packet is not negligible only in the region
having the volume of the order of $D^3$ and so if ${\bf r}_1$ is inside this region then
the probability to detect the particle is of the order of $(d/D)^3$.

If the particle is detected by the device then the measured momentum range is defined
by the Fourier transform of ${\tilde\psi}({\bf r},t)$. A direct calculation gives
\begin{eqnarray}
&&{\tilde\psi({\bf p},t)}=\frac{1}{(2\pi\hbar)^{3/2}}\int exp(\frac{-i}{\hbar}{\bf p}{\bf r})
{\tilde\psi({\bf r},t)}d^3{\bf r}=\nonumber\\
&&f({\bf p},t)exp[-\frac{d^2D^2a^2({\bf p}-m{\bf r}_1/t)^2}
{2\hbar^2(D^2a^2+d^4)}]
\label{star}
\end{eqnarray} 
where $f({\bf p},t)$ contains the dependence on ${\bf p}$ only in the exponent with the
imaginary index. Therefore the probabilities of different momenta are defined by the last
exponent which shows that the distribution of momenta has a sharp peak around the
vector $m{\bf r}_1/t$ pointing to the device. While the width of the momentum 
distribution in the initial packet is of order of
${\hbar}/a$ (see Eq. (\ref{chipt})), the width given by Eq. (\ref{star}) is much narrower.
If for example $D^2a^2\gg d^4$ then the width is of the of the order of ${\hbar}/d$
and in the opposite case  the width is of the of the order of ${\hbar}d/(Da)$.

As discussed in Sec. \ref{classical}, in semiclassical approximation the value of the 
momentum can be found by applying the
operation $-i\hbar \partial/\partial{\bf r}$ to the rapidly oscillating exponent. In general the
momentum distribution can be rather wide. However, if the particle is detected in a
vicinity of the point ${\bf r}$ then, as follows from Eq. (\ref{star}),
it will be detected with the momentum close to
$m{\bf r}/t$. This result has the following qualitative explanation. The operation
$-i\hbar \partial/\partial{\bf r}$ applied to the imaginary index of the exponent in Eq. 
(\ref{psirt}) gives exactly $m{\bf r}/t$.

The above results gives the solution of the Mott-Heisenberg problem when
the particle is in the state (\ref{psirt}). However, in this case the wave function can be
spherically symmetric only if ${\bf p}_0=0$. This case is of no interest because typically
a particle created in the spherically symmetric state has a nonzero kinetic energy.
We now consider a model where, instead of Eq. (\ref{chip}), the initial particle
momentum wave function is
\begin{equation}
\chi({\bf p})=\frac{f({\bf p}/p)}{p}exp[-\frac{1}{2\hbar^2}a^2(p-p_0)^2]
\label{chipn}
\end{equation}
where $p=|{\bf p}|$ and the quantaties $p_0$ and $a$ are 
positive. We assume that $p_0a\gg \hbar$. Then, with a good accuracy, integrals
over $p$ from $0$ to $\infty$ containing the exponent can be replaced by
integrals from $-\infty$ to $\infty$.  By analogy with the calculation in Sec. \ref{classical}, 
one can easily show that ${\bar p}\approx p_0$ and $\Delta p\approx \hbar/(a\sqrt{2})$
and therefore  the $p$-distribution is semiclassical. The dependence of
the momentum wave function on $t$ is the same as in Eq. (\ref{chipt}).

The coordinate wave function is again given by Eq. (\ref{Fourier}). For calculating this
function in the case when the initial momentum wave function is given by Eq. (\ref{chipn})
we need the following auxiliary results:  
\begin{equation} 
\int_0^{\infty} exp[-\frac{1}{2\hbar^2}a^2(p-p_0)^2+\frac{i}{\hbar}pr \xi]dp\approx 
\frac{\hbar}{a}(\frac{2\pi}{1+iD/a})^{1/2}exp[-\frac{(r\xi-p_0t/m)^2}{2a^2(1+iD/a)}]
\label{rxi}
\end{equation}
where $r=|{\bf r}|$ and
\begin{equation}
exp(\frac{i}{\hbar}{\bf pr})=4\pi\sum_{l\mu}i^lj_l(pr/\hbar )Y_{l\mu}^*({\bf p}/p)Y_{l\mu}({\bf r}/r)
\label{flat}
\end{equation}
The last expression is the well-known decomposition of the flat wave. Here $Y_{l\mu}$ is the
spherical function corresponding to the orbital angular momentum $l$ and its $z$-projection $\mu$
and $j_l$ is the spherical Bessel function. Its asymptotic expression when the argument is large
is $j_l(x)\approx sin(x-\pi l/2)/x$.

Let $f({\bf p}/p)=\sum_{l\mu}c_{lm}Y_{l\mu}({\bf p}/p)$ be the decomposition of the
function $f$ in Eq. (\ref{chipn}) over spherical functions. Then it follows from the orthogonality
of spherical functions, Eqs. (\ref{chipn}-\ref{flat}) and the above remarks that
if $(pr/\hbar)\gg 1$ then
\begin{eqnarray}
&&\psi({\bf r},t)=-\frac{i}{ar}(\frac{\hbar}{1+iD/a})^{1/2}exp(-\frac{p_0^2t}{2m\hbar})
\sum_{l\mu}c_{l\mu}Y_{l\mu}({\bf r}/r)\nonumber\\
&&\{exp[-\frac{(r-p_0t/m)^2}{2a^2(1+iD/a)}]-(-1)^lexp[-\frac{(r+p_0t/m)^2}{2a^2(1+iD/a)}]\}
\label{psirn}
\end{eqnarray}
At large distances and times the second term in the figure brackets is negligible and the final
result is
\begin{eqnarray}
\psi({\bf r},t)=-\frac{i}{ar}(\frac{\hbar}{1+iD/a})^{1/2}exp(-\frac{p_0^2t}{2m\hbar})
f({\bf r}/r)exp[-\frac{(r-p_0t/m)^2(1-iD/a)}{2(a^2+D^2)}]
\label{psif}
\end{eqnarray}

{\it Therefore for the initial momentum wave function (\ref{chipn}) the coordinate wave function at large distances and 
times has the same angular dependence as the momentum wave
function and the radial wave functions spreads out by analogy with Eq. (\ref{psirt}).}

The result (\ref{psif}) gives an obvious solution of the Mott-Heisenberg problem in the case
when the angular dependence of the wave function is arbitrary. Indeed, suppose that a particle
is created at the origin and a measuring device is seen from the origin in the narrow angular range
defined by the function ${\tilde f}({\bf r}/r)$. Suppose that the support of ${\tilde f}({\bf r}/r)$
is within the range defined by $f({\bf r}/r)$. Then the projection of the wave function (\ref{psif})
onto the device is given by the same expression where $f({\bf r}/r)$ is replaced by 
${\tilde f}({\bf r}/r)$. Since the angular wave functions in coordinate and momentum 
representations are the same, the momenta  measured by the device will be in the angular range
defined by the function ${\tilde f}({\bf p}/p)$.

\section{Position operator in relativistic quantum mechanics}
\label{momentum}

The problem of the position operator in relativistic quantum theory has been discussed in a wide literature
and different authors have different opinions on this problem. In particular, some authors state that in
relativistic quantum theory no position operator exists. As already noted, the results of fundamental quantum theories
are formulated only in terms of the S-matrix in momentum space without any mentioning of space-time. This is in
the spirit of the Heisenberg S-matrix program that in relativistic quantum theory it is possible to describe only
transitions of states from the infinite past when
$t\to -\infty$ to the distant future when $t\to +\infty$. On the other hand, since quantum theory is treated as a
theory more general than classical one, it is not possible to fully avoid space and time in quantum theory.
For example, quantum theory should explain how photons from distant objects travel to Earth and even  
how macroscopic bodies are moving along classical trajectories. Hence we can conclude that: 
a) in quantum theory (nonrelativistic and relativistic) we must have a position operator and 
b) this operator has a physical meaning only in semiclassical approximation.

Let us first consider the definition of elementary particle. 
Although theory of elementary particles exists for a rather long period of time, there is no commonly accepted definition of elementary particle in this theory. 
In Refs. \cite{symm1401,DS} and references therein we argue that, in the spirit of Wigner's approach to Poincare symmetry
\cite{Wigner}, a general definition, not depending on the choice of the classical background 
and on whether we consider a local or nonlocal theory, is  that a particle is elementary if 
the set of its wave functions is the space of an IR of the symmetry algebra in the given theory. 

There exists a wide literature describing how IRs of the Poincare algebra
can be constructed. In particular, an IR for a spinless particle can be implemented in a space of functions $\xi({\bf p})$ 
satisfying the condition
\begin{equation}
\int |\xi({\bf p})|^2d\rho({\bf p})<\infty, \quad d\rho({\bf p})=\frac{d^3{\bf p}}{\epsilon({\bf p})}
\label{invnorm}
\end{equation}
where $\epsilon({\bf p})=(m^2+{\bf p}^2)^{1/2}$ is the energy of the particle with the mass $m$. The convenience
of the above requirement is that the volume element $d\rho({\bf p})$ is Lorentz invariant. In that case 
it can be easily shown by direct calculations (see e.g. Ref. \cite{current}) that the representation operators 
have the form
\begin{eqnarray}
{\bf L}=-i{\bf p}\times\frac{\partial}{\partial {\bf p}},\quad 
{\bf N}=-i\epsilon({\bf p})\frac{\partial}{\partial {\bf p}}, \quad {\bf P}={\bf p},\quad E=\epsilon({\bf p})
\label{IRoperators}
\end{eqnarray}
where ${\bf L}$ is the orbital angular momentum operator, ${\bf N}$ is the Lorentz
boost operator, ${\bf P}$ is the momentum operator, $E$ is the energy operator and these operators are expressed
in terms of the operators in Eq. (\ref{PCR}) as
$${\bf L}=(M^{23},M^{31},M^{12}),\,\, {\bf N}=(M^{10},M^{20},M^{30}),\,\,
{\bf P}=(P^1,P^2,P^3),\,\,E=P^0$$

For particles with spin these results are modified as follows. For a massive particle with spin $s$ the 
functions $\xi({\bf p})$ also depend on spin projections which can take $2s+1$ values $-s,-s+1,...s$. If ${\bf s}$ is the spin operator then the total angular momentum has an additional term ${\bf s}$ and the Lorentz boost operator has 
an additional term $({\bf s}\times{\bf p})/(\epsilon({\bf p})+m)$ (see e.g. Eq. (2.5) in Ref. \cite{current}). Hence corrections of the spin terms to the quantum 
numbers describing the angular momentum and the Lorentz boost do not exceed $s$. We assume as usual that in semiclassical approximation the quantum numbers characterizing the angular momentum and the Lorentz boost are much greater than
unity and hence in this approximation spin effects can be neglected. For a massless particle with the spin $s$ the
spin projections can take only values $-s$ and $s$ and those quantum numbers have the meaning of helicity. In this
case the results for the representation operators can be obtained by taking the limit $m\to 0$ if the operators
are written in the light front variables (see e.g. Eq. (25) in Ref. \cite{symm1401}). As a consequence,
in semiclassical approximation the spin corrections in the massless case can be neglected as well. Hence for investigating
the position operator we will neglect spin effects and will not explicitly write the dependence of wave functions 
on spin projections. 

In the above IRs the representation operators are Hermitian as it should be for operators corresponding to physical
quantities. In standard theory (over complex numbers) such IRs of the Lie algebra can be extended to unitary IRs
of the Poincare group. In particular, in the spinless case the unitary operator $U(\Lambda)$ corresponding to the
Lorentz transformation $\Lambda$ acts in $H$ as (see e.g. Ref. \cite{current})
\begin{equation}
U(\Lambda)\xi(p)=\xi(\Lambda^{-1}p)
\label{U(Lambda)}
\end{equation}

In the literature elementary particles are described not only by such IRs but
also by nonunitary representations induced from the Lorentz group \cite{Mensky}.
Since the factor space of the Poincare group over the Lorentz group is Minkowski space, the elements of such 
representations are fields $\Psi(x)$ depending on four-vectors $x$ in Minkowski space and possibly on spin variables.
Since those functions describe nonunitary representations, their probabilistic interpretation is problematic.
The Pauli theorem \cite{Pauli2} states that for fields with an integer spin it is impossible to define a
positive definite charge density and for fields with a half-integer spin it is impossible to define a
positive definite energy density. 

Hence a problem arises whether such fields have a physical meaning.
The answer is that in QFT after quantizing they become quantum fields defining the stress-energy and angular momentum
tensors. Then the Hermitian operators $P^{\mu}$ and $M^{\mu\nu}$ are defined by integrals of those tensors
over a space-like hyperplane. So the quantity $x$ in local fields is only an integration parameter and 
a problem of whether there are quantum operators corresponding to $x$ does not arise. This is clear also from the fact 
that quantized fields are operators in Fock spaces describing systems with an infinite number of particles and
hence $x$ does not refer to any specific particle. Therefore local 
quantum fields (in this situation even the term
"local" is not clear) are only auxiliary tools for constructing physical operators in QFT.

Let us note that although QFT has achieved very impressive successes in explaining many experimental data,
a problem of its mathematical substantiation has not been solved yet. The main mathematical inconsistency of QFT is that
since interacting local quantum fields can be treated only as operatorial distributions, their products at the
same space-time points are not well defined (see e.g. Ref. \cite{Bogolubov}). One of ideas of the string 
theory is that if products of fields at the same points (zero-dimensional objects) are replaced by products where
arguments belong to strings (one-dimensional objects) then there is 
hope that infinities will be less singular. In view of such controversial properties of local quantum fields,
many authors posed a question whether local fields will survive in the future quantum theory. 
Nevertheless, in the literature
the problem of position operator is mainly discussed in the approach when elementary particles are described
by local fields rather than unitary IRs. Below we discuss the both approaches 
but first we consider the case of unitary IRs.

As follows from Eq. (\ref{PCR}), the operator $I_2=E^2-{\bf P}^2$ is the Casimir operator 
of the second order, 
i.e. it is a bilinear combination of representation operators commuting with
all the operators of the algebra. As follows from the well-known Schur lemma, all states belonging to an IR are
the eigenvectors of $I_2$ with the same eigenvalue $m^2$. Note that Eq. (\ref{IRoperators}) contains only $m^2$
but not $m$. The choice of the energy sign is only a matter of convention
but not a matter of principle. Indeed, the energy can be measured only if the momentum ${\bf p}$ is measured
and then it is only a matter of convention what sign of the square root should be chosen. However, it is
important that the sign should be the same for all particles. For example, if we consider a system of two
particles with the same values of $m^2$ and the opposite momenta ${\bf p}_1$ and ${\bf p}_2$ such that
${\bf p}_1+{\bf p}_2=0$, we cannot define the energies of the particles as $\epsilon({\bf p}_1)$
and $-\epsilon({\bf p}_2)$, respectively, since in that case the total four-momentum of the two-particle
system will be zero what contradicts experiment.

The notation $I_2=m^2$ is justified by the fact that for all known particles $I_2\geq 0$. Then the mass 
$m$ is {\it defined} as the square root of $m^2$ and the sign of $m$ is only a
matter of convention. The usual convention is that $m\geq 0$. However, from mathematical point of view, 
IRs with $I_2<0$ are not prohibited. If the velocity
operator ${\bf v}$ is {\it defined} as ${\bf v}={\bf P}/E$ then for known particles $|{\bf v}|\leq 1$, i.e.
$|{\bf v}|\leq c$ in standard units. However, for IRs with $I_2 < 0$, $|{\bf v}|> c$ and,
at least from the point of view of mathematical construction of IRs, this case is not prohibited. The hypothetical
particles with such properties are called tachyons and their possible existence is widely discussed in
the literature. If the tachyon mass $m$ is also defined as the square root of $m^2$ then this quantity will
be imaginary. However, this does not mean than the corresponding IRs are unphysical since all the operators
of the Poincare group Lie algebra depend only on $m^2$.

As follows from Eqs. (\ref{invnorm}) and (\ref{IRoperators}), in the nonrelativistic approximation 
$d\rho({\bf p})=d^3{\bf p}/m$ and ${\bf N}=-im\partial /\partial{\bf p}$. Therefore in this approximation ${\bf N}$
is proportional to {\it standard} position operator and one can say that the position operator is
in fact present in the description of the IR.

The following remarks are in order. The choice of the volume element in the Lorentz invariant 
form $d\rho({\bf p})$ (see Eq. (\ref{invnorm})) might be convenient from the point of view that then
the Hilbert space can be treated as a space of functions $\xi(p)$ depending on four-vectors $p$ such that
$p^0=\epsilon({\bf p})$ and the norm can be written in the covariant form (i.e. in the form depending only on
Lorentz invariant quantities):
$||\xi||^2=\int |\xi(p)|^2\delta (p^2-m^2)\theta (p^0)d^4p$. However, the requirement of covariance does not have
a fundamental physical meaning. In relativistic theory a necessary requirement is that symmetry is defined by
operators satisfying the commutation relations (\ref{PCR}) and this requirement can be implemented in different 
forms, not necessarily in covariant ones. 

As an illustration, consider the following problem. Suppose that we wish to construct a single-particle coordinate
wave function. Such a wave function cannot be defined on the whole Minkowski space. This is clear even from the
fact that there is no time operator. The wave function can be defined only on a space-like hyperplane of the
Minkowski space. For example, on the hyperplane $t=const$ the wave function depends only on ${\bf x}$. Hence
for defining the wave function one has to choose the form of the position operator. By analogy with the
nonrelativistic case, one might try to define the position operator as $i\partial/\partial{\bf p}$. However,
if the Hilbert space is implemented as in Eq. (\ref{invnorm}) then this operator is not selfadjoint since $d\rho({\bf p})$
is not proportional to $d^3{\bf p}$. One can perform a unitary transformation 
$\xi({\bf p})\to \chi({\bf p})=\xi({\bf p})/\epsilon({\bf p})^{1/2}$ such that the Hilbert space becomes
the space of functions $\chi({\bf p})$ satisfying the condition $\int|\chi({\bf p})|^2d^3{\bf p}<\infty$. 
It is easy to verify that in this implementation of the IR the operators $({\bf L},{\bf P},E)$ will have
the same form as in Eq. (\ref{IRoperators}) but the expression for ${\bf N}$ will be
\begin{equation}
{\bf N}=-i\epsilon({\bf p})^{1/2}\frac{\partial}{\partial {\bf p}}\epsilon({\bf p})^{1/2}
\label{newN}
\end{equation}
In this case one can {\it define} $i\hbar \partial/\partial{\bf p}$ as a position operator
but now we do not have a situation when the position operator is present among the other representation
operators. 

A problem of the definition of the position operator in relativistic quantum theory has been discussed since
the beginning of the 1930s and it has been noted that when quantum theory is combined with relativity the
existence of the position operator with correct physical properties becomes a problem. The above definition
has been proposed by Newton and Wigner in Ref. \cite{NW}. They worked in the approach when elementary particles
are described by local fields $\Psi(x)$ defined on the whole Minkowski space rather than unitary IRs. As
noted above, such fields cannot be treated as single-particle wave functions. The spacial Fourier transform 
of such fields at $t=const$ describes states
where the energy can be positive and negative and this is interpreted such that local quantum fields describe
a particle and its antiparticle simultaneously. Newton and Wigner first discuss the spinless case and consider
only states on the upper Lorentz hyperboloid where the energy is positive. For such states the representation
operators act in the same way as in the case of spinless unitary IRs. With this definition 
the coordinate wave function $\psi({\bf r})$ can be again defined by Eq. (\ref{psir}) and a question arises 
whether such a position operator has all the required properties.

For example, in the introductory section of the
textbook \cite{BLP} the following arguments are given in favor of the statement that in relativistic quantum theory 
it is not possible to define a physical position operator.
Suppose that we measure coordinates of an electron with the mass $m$. When the uncertainty of  
coordinates is of the order of $\hbar/mc$, the uncertainty of momenta is of the order of $mc$, the uncertainty
of energy is of the order of $mc^2$ and hence creation of electron-positron pairs is allowed. As a consequence,
it is not possible to localize the electron with the accuracy better than its Compton wave length
${\hbar}/mc$. Hence, for a particle with a nonzero mass exact measurement is possible only
either in the nonrelativistic limit (when $c\to\infty$) or
classical limit (when ${\hbar}\to 0)$. In the case of the photon, as noted by Pauli (see p. 191 of Ref.
\cite{Pauli}), the coordinate cannot be measured with the accuracy better than $\hbar/p$ where $p$ is the magnitude
of the photon momentum. The quantity $\lambda=2\pi\hbar/p$ is called the photon wave length although, as noted
in Sec. \ref{intropos}, the meaning of this quantity in quantum case might be fully different than in classical one. Since
$\lambda\to 0$ in the formal limit $\hbar\to 0$, Pauli concludes that "Only within the confines of the classical ray
concept does the position of the photon have a physical significance".

Another argument that the Newton-Wigner position operator does not have all the required properties follows.
Since the energy operator acts on the function $\chi({\bf p})$ as $E\chi({\bf p})=\epsilon({\bf p})\chi({\bf p})$
(see Eq. (\ref{IRoperators})) and the energy is an operator corresponding to infinitesimal time translations,
the dependence of the wave function $\chi({\bf p})$ on $t$ is given by 
\begin{equation}
\chi({\bf p},t)=exp(-\frac{i}{\hbar}Et)\chi({\bf p})=exp(-\frac{i}{\hbar}\epsilon({\bf p})t)\chi({\bf p})
\label{chiptA}
\end{equation}
Then a relativistic analog of Eq. (\ref{tnonrel}) is 
\begin{equation}
\psi({\bf r},t)=\int exp\{\frac{i}{\hbar}[{\bf p}({\bf r}-{\bf r}')-\epsilon({\bf p})t]\}\psi_0({\bf r}')
\frac{d^3{\bf r}'d^3{\bf p}}{(2\pi\hbar)^3}
\label{trel}
\end{equation}
As a consequence, the Newton-Wigner position operator has the "tail property": if $\psi_0({\bf r})\neq 0$ only if ${\bf r}$ belongs to a finite vicinity of some vector ${\bf r}_0$ then at any $t>0$ the function $\psi({\bf r},t)$ 
has a tail belonging to the whole three-dimensional space, i.e. the
wave function spreads out with an infinite speed. Hence at any $t>0$ the
particle can be detected at any point of the space and this contradicts the requirement that no information
should be transmitted with the speed greater than $c$. 

The tail property of the Newton-Wigner position operator has been known for a long time (see e.g. Ref. \cite{Hegerfeldt}
and references therein). It is characterized as nonlocality leading to the action at a distance. Hegerfeldt argues \cite{Hegerfeldt} that this property is rather general because it can be proved assuming that energy is positive and without assuming a specific choice of the position operator. 
The Hegerfeldt theorem \cite{Hegerfeldt} is based on the assumption that there exists an operator $N(V)$
whose expectation defines the probability to find a particle inside the volume $V$. However, 
the meaning of time on quantum level is not clear and for the position operator
proposed in the present paper such a probability does not exist because there is no wave function in coordinate
representation (see Sec. \ref{consistent} and the discussion in Sec. \ref{conclusion}). 

One might say that the requirement that no signal can be transmitted with the speed greater than $c$ has been obtained in
Special Relativity which is a classical (i.e. nonquantum) theory operating only with classical space-time
coordinates. For example, in classical theory the velocity of a particle is defined as ${\bf v}=d{\bf r}/dt$
but, as noted above, the velocity {\it should be defined} as ${\bf v}={\bf p}/E$ (i.e. without mentioning 
space-time) and then on classical level it can be shown that ${\bf v}=d{\bf r}/dt$. 
In QFT local quantum fields separated by space-like intervals commute or anticommute
(depending on whether the spin is integer or half-integer) and this is treated as a requirement of causality
and that no signal can be transmitted with the speed greater than $c$. However, as noted above, the physical
meaning of space-time coordinates on quantum level is not clear. Hence from the point of view of quantum theory 
the existence of tachyons is not prohibited. Note also that when two electrically charged particles exchange by a virtual photon, a typical situation is that the four-momentum of the photon is space-like, i.e. the photon is the tachyon. 
We conclude that although in relativistic theory such a behavior might seem undesirable, there is no proof that it must 
be excluded. Also, as argued by Griffiths (see Ref. \cite{Griffiths} and references therein), with a consistent 
interpretation of quantum theory there are no nonlocality and superluminal interactions. In Sec. \ref{conclusion} we argue that the position operator proposed in the present paper sheds a new light on this problem.

Another striking example is a photon emitted in the famous
21cm transition line between the hyperfine energy levels of the hydrogen atom. The phrase that the lifetime of 
this transition is of the order of $\tau=10^7$ years implies that the width of the level is of the order of $\hbar/\tau$, 
i.e. experimentally the uncertainty of the photon energy is $\hbar/\tau$. Hence the uncertainty of the
photon momentum is $\hbar/(c\tau)$ and with the above definition of the coordinate operators the uncertainty of the longitudinal coordinate is $c\tau$, i.e. of the order of $10^7$ light years. Then there is a nonzero probability that 
immediately after its creation at point A the photon can be detected at point B such that the distance between A 
and B is $10^7$ light years. 

A problem arises how this phenomenon should be interpreted. On one hand, one might say that in view of the above
discussion it is not clear whether or not the requirement that no information should be transmitted with the speed greater than $c$ should be a must in relativistic quantum theory. On the other hand (as pointed out to me by Alik Makarov), 
we can know about the photon creation only if the photon is detected and when it 
was detected at point B at the moment of time $t=t_0$, this does not mean that the photon 
travelled from A to B with the speed greater than $c$ since the time of creation has an uncertainty of the order 
of $10^7$ years. Note also that in this situation a description of the system (atom + electric
field) by the wave function (e.g. in the Fock space) depending on a continuous parameter $t$ has no physical meaning 
(since roughly speaking the quantum of time in this process is of the order of $10^7$ years). 
If we accept this explanation then we should acknowledge that in some situations a description of
evolution by a continuous classical parameter $t$ is not physical and this is in the spirit of the Heisenberg S-matrix
program. However, this example describes a pure quantum phenomenon while, as noted above, a position operator is
needed only in semiclassical approximation.

For particles with nonzero spin, the number of states in local fields is typically by a factor of two greater
than in the case of unitary IRs (since local fields describe a particle and its antiparticle simultaneously) but those
components are not independent since local fields satisfy a covariant equation (Klein-Gordon, Dirac etc.). In Ref.
\cite{NW} Newton and Wigner construct a position operator in the massive case but say that in the massless one they
have succeeded in constructing such an operator only for Klein-Gordon and Dirac particles while in the case of the
photon the position operator does not exist. On the other hand, as noted above, in the case of unitary IRs different
spin components are independent and in semiclassical approximation spin effects are not important. So in this
approach one might adopt the Newton-Wigner position operator for particles with any spin and any mass. 

We now consider the following problem. Since the Newton-Wigner position operator formally has the same form as in
nonrelativistic quantum mechanics, the coordinate and momentum wave functions also are related to each other by the same 
Fourier transform as in nonrelativistic quantum mechanics (see Eq. (\ref{Fourier})). One might think that
this relation is not Lorentz covariant  and pose a question whether in relativistic theory this is acceptable. As noted above, for constructing 
the momentum wave function covariance does not have a fundamental physical meaning and is not necessary. A question
arises whether the same is true for constructing the coordinate wave function.

Let us note first that if the four-vector $x$ is such that $x=(t,{\bf x})$ then the wave function 
$\psi(x)=\psi({\bf x},t)$ can have a physical meaning only if we accept that (at least in some approximations) a
position operator is well defined. Then the function $\psi({\bf x},t)$ describes amplitudes of probabilities for
different values of ${\bf x}$ at a fixed value of $t$. This function cannot describe amplitudes of probabilities for
different values of $t$ because there is no time operator. 

For discussing Lorentz covariance of the coordinate wave function it is important to note that, in view of the above
remarks, this function can be defined not in the whole Minkowski space but only on space-like hyperplanes of that
space (by analogy with the fact that in QFT the operators $(P^{\mu},M^{\mu\nu})$ are defined by integrals
over such hyperplanes). They are defined by a time-like unit vector $n$ and the evolution parameter $\tau$ such that 
the corresponding hyperplane is a set of points with the coordinates $x$ satisfying the condition $nx=\tau$.
Wave functions $\psi(x)$ on this hyperplane satisfy the requirement $\int |\psi(x)|^2\delta(nx-\tau)d^4x<\infty$. In a
special case when $n^0=1$, ${\bf n}=0$ the hyperplane is a set of points $(t=\tau,{\bf x})$ and the wave
functions satisfy the usual requirement $\int |\psi({\bf x},t)|^2d^3{\bf x}<\infty$. In the literature coordinate wave functions are usually considered without discussions of the position operator
and without mentioning the fact that those functions are defined on space-like hyperplanes (see e.g. Refs.
\cite{SR,Bradler}).

By analogy with the construction of the coordinate wave function in Refs.  \cite{SR,Naumov}, it can be defined as
follows. Let ${\tilde x}_0$ be a four-vector and $p$ and $p_0$ be four-vectors
$(\epsilon({\bf p}),{\bf p})$ and $(\epsilon({\bf p}_0),{\bf p}_0)$, respectively. We will see below that momentum wave functions describing wave packets can be chosen in the form
\begin{equation}
\xi(p,p_0,{\tilde x}_0)=f(p,p_0)exp(\frac{i}{\hbar}p{\tilde x}_0)
\label{covxi}
\end{equation}
where $f(p,p_0)$ as a function of $p$ has a sharp maximum in the vicinity of $p=p_0$, ${\tilde x}_0=x_0-(nx_0)n$ and
the four-vector $x_0$ has the coordinates $(t,{\bf r}_0)$. Then the coordinate wave function can be defined as
\begin{equation}
\psi(x,p_0,{\tilde x}_0)=\frac{1}{(2\pi\hbar)^{3/2}}\int \xi(p,p_0,{\tilde x}_0)exp(-\frac{i}{\hbar}px)d\rho({\bf p})
\label{covpsi}
\end{equation}
Suppose that $f(p,p_0)$ is a covariant function of its arguments, i.e. it can depend only on $p^2$, $p_0^2$ and
$pp_0$. Then, as follows from Eq. (\ref{U(Lambda)}), the function $\psi(x,p_0,{\tilde x}_0)$ is covariant because 
its Lorentz 
transformation is $\psi(x,p_0,{\tilde x}_0)\to \psi(\Lambda^{-1}x, p_0, {\tilde x}_0)$.

The choice of $f(p,p_0)$ in the covariant form might encounter the following problem. For example, the authors of
Ref. \cite{Naumov} propose to consider $f(p,p_0)$ in the form
\begin{equation}
f(p,p_0)=const\, exp[\frac{(p-p_0)^2}{4\sigma^2}]
\label{Naumov}
\end{equation}
The exponent in this expression has the maximum at ${\bf p}={\bf p}_0$ and in the vicinity of the maximum
\begin{equation}
(p-p_0)^2=-({\bf p}-{\bf p}_0)^2+[\frac{({\bf p}_0,{\bf p}-{\bf p}_0)}{\epsilon({\bf p}_0)}]^2+o(|{\bf p}-{\bf p}_0|^2)
\label{p-p0}
\end{equation}
If ${\bf p}_0$ is directed along the $z$ axis and the subscript ${\bot}$ is used to denote the projection of
the vector onto the $xy$ plane then
\begin{equation}
(p-p_0)^2=-({\bf p}_{\bot}-{\bf p}_{0\bot})^2-[\frac{m}{\epsilon({\bf p}_0)}]^2(p_z-p_{0z})^2+o(|{\bf p}-{\bf p}_0|^2)
\label{p-p0B}
\end{equation}
It follows from this expression that if the particle is ultrarelativistic then the width of the momentum distribution
in the longitudinal direction is much greater that in transverse ones and for massless particles the former becomes
infinite. We conclude that for massless particles the covariant parametrization of $f(p,p_0)$ is problematic.

As noted above, the only fundamental requirement on quantum level is that the representation operators should satisfy
the commutation relations (\ref{PCR}) while covariance is not fundamental. Nevertheless, the above discussion shows
that covariance of coordinate wave functions can be preserved if one takes into account the fact that they are 
defined on space-like hyperplanes. In particular, covariance of functions $f$ can be preserved if one assumes
that they depend not only on $p$ and $p_0$ but also on $n$. In what follows we consider only the case when the vector 
$n$ is such that $n^0=1$ and ${\bf n}=0$. Let us replace $f(p,p_0)$ by $f({\tilde p},{\tilde p}_0)$ where ${\tilde p}=p-(pn)n$ and ${\tilde p}_0=p_0-(p_0n)n$. Then the four-vectors ${\tilde p}$ and ${\tilde p}_0$ have only nonzero
spatial components equal ${\bf p}$ and ${\bf p}_0$, respectively. As a consequence, any rotationally invariant 
combination of 
${\bf p}$ and ${\bf p}_0$ can be treated as a Lorentz covariant combination of ${\tilde p}$ and ${\tilde p}_0$.

We conclude that with the above choice of the vector $n$ one can work with momentum and coordinate wave functions
in full analogy with nonrelativistic quantum mechanics and in that case Lorentz covariance is satisfied. In
particular in that case Eq. (\ref{covpsi}) can be written in the form of Eq. (\ref{Fourier}).

We now consider the photon case in greater details. The coordinate photon wave function has been discussed by
many authors. A question arises in what situations this function is needed. As already noted, since the
fundamental theory of electromagnetic interactions is QED, and this theory does not 
contain space-time at all,
for solving quantum problems in the framework of QED the coordinate photon wave function is not needed.
However, this function is used in some special problems, for example for describing single-photon interference
and diffraction by analogy with classical theory.

In the present paper we consider only the case of free photons. If we consider a motion of a free particle, 
it is not important in what interactions this particle
participates and, as explained above, if the particle is described by its IR in semiclassical approximation then the particle spin is not important. Hence the effect of WPS for an ultrarelativistic particle does not depend on
the nature of the particle, i.e. on whether the particle is the photon, the proton, the electron etc. 
For this reason we are interested in papers on the photon coordinate wave function mainly from the point
of view how the position operator for the free ultrarelativistic particle is defined.

Note that in classical theory the notion of field, as well as that of wave, is used for describing systems of many particles 
by their  mean characteristics. For example, the electromagnetic field consists of many photons. In classical theory each
photon is not described individually but the field as a whole is described by the field strengths ${\bf E}({\bf r},t)$ and 
${\bf B}({\bf r},t)$ which can be measured (in principle) by using macroscopic test bodies
such that the quantities ${\bf r}$ and $t$ refer to positions of such bodies at time $t$. In quantum theory one can
formally define corresponding quantized field operators but the meaning of $({\bf r},t)$ for elementary particles
is not clear. In addition, in view of the above remarks, the physical meaning of electric and magnetic fields of a free photon is problematic. 

For the first time the coordinate photon wave function has been discussed by Landau and Peierls in Ref. \cite{LP}.
However, in the literature it has been stated (see e.g. Refs. \cite{AB} and \cite{SR}) that in QED there is no way 
to define a coordinate photon wave function. A section in the textbook \cite{AB} is titled "Impossibility of introducing the
photon wave function in coordinate representation". The arguments follow. The electric
and magnetic fields of the photon in coordinate representation are proportional to the Fourier transforms
of $|{\bf p}|^{1/2}\chi({\bf p})$, rather than $\chi({\bf p})$. As a consequence, the quantities ${\bf E}({\bf r})$
and ${\bf B}({\bf r})$ are defined not by $\psi({\bf r})$ but by integrals of $\psi({\bf r})$ over a 
region of the order of the wave length. However, this argument also does not exclude the possibility that 
$\psi({\bf r})$ can have
a physical meaning in semiclassical approximation since, as noted above, the notions of the electric and
magnetic fields of a single photon are problematic. In addition, since $\lambda\to 0$ in the
formal limit $\hbar\to 0$, one should not expect that any position operator in semiclassical approximation
can describe coordinates with the accuracy better than the wave length. Another arguments in favor
of the existence of the coordinate photon wave function have been given by Bialynicki-Birula \cite{Bialy}. 

A detailed discussion of the photon position operator can be found in papers by Margaret Hawton and
references therein (see e.g. Ref. \cite{Hawton}). In this approach the photon is described by a local field
and the momentum and coordinate representations are related to each other by standard Fourier transform.
The author of Ref. \cite{Hawton} discusses generalizations of the photon position operator proposed by
Pryce \cite{Pryce}. However, the Pryce operator and its generalizations discussed in Refs. \cite{Bialy,Hawton}
differ from the Newton-Wigner operator only by terms of the order of the wave length. Hence in semiclassical approximation all those operators are equivalent. 

The above discussion shows that on quantum level the physical meaning of the coordinate is a difficult problem but
in view of a) and b) (see the beginning of this section) one can conclude that in semiclassical
approximation all the existing proposals for the position operator are equivalent to the  
Newton-Wigner operator $i\hbar \partial/\partial{\bf p}$. An additional argument in favor of this operator
is that the relativistic nature of the photon might be somehow manifested in the longitudinal direction while
in transverse directions the behavior of the wave function should be similar to that in standard
nonrelativistic quantum mechanics. Another argument is that the photon wave function in coordinate 
representation constructed by using this
operator satisfies the wave equation in agreement with classical electrodynamics (see Sec. \ref{geom}).

For all the reasons described above, in the next section we consider what happens if the 
space-time evolution of relativistic wave packets is described by using the Newton-Wigner position operator.

\section{Wave packet spreading in relativistic quantum mechanics}
\label{RelWPS}

Consider first a construction of the wave packet for a particle with nonzero mass. A possible way of the
construction follows. We first consider the particle in its rest system, i.e. in the reference frame where
the mean value of the particle momentum is zero. The wave function $\chi_0({\bf p})$ in this case can be taken
as in Eq. (\ref{chip}) with ${\bf p}_0=0$. As noted in Sec. \ref{classical}, such a state cannot be 
semiclassical. However, it is possible to obtain a semiclassical state by applying a Lorentz transformation
to $\chi_0({\bf p})$. As a consequence of Eq. (\ref{U(Lambda)}) and the relation between the functions $\xi$ and
$\chi$ 
\begin{equation}
U(\Lambda)\chi_0({\bf p})=[\frac{\epsilon({\bf p}')}{\epsilon({\bf p})}]^{1/2}\chi_0({\bf p}')
\label{Lorentz}
\end{equation} 
where ${\bf p}'$ is the momentum obtained from ${\bf p}$ by the Lorentz transformation $\Lambda^{-1}$. If $\Lambda$ is
the Lorentz boost along the $z$ axis with the velocity $v$ then
\begin{equation}
{\bf p}_{\bot}'={\bf p}_{\bot},\quad p_z'=\frac{p_z-v\epsilon({\bf p})}{(1-v^2)^{1/2}}
\label{p'}
\end{equation}

As follows from this expression, $exp(-{\bf p}^{'2}a^2/2\hbar^2)$ as a function of ${\bf p}$ has the maximum
at ${\bf p}_{\bot}=0$, $p_z=p_{z0}=v[(m^2+{\bf p}_{\bot}^2)/(1-v^2)]^{1/2}$ and near the maximum
$$exp(-\frac{a^2{\bf p}^{'2}}{2\hbar^2})\approx exp\{-\frac{1}{2\hbar^2}[a^2{\bf p}_{\bot}^2+b^2(p_z-p_{z0})^2]\}$$
where $b=a(1-v^2)^{1/2}$ what represents the effect of the Lorentz contraction. If $mv\gg \hbar/a$ (in units where $c=1$) 
then $m\gg |{\bf p}_{\bot}|$  and $p_{z0}\approx mv/(1-v^2)^{1/2}$. In this case the transformed state is semiclassical
and the mean value of the momentum is exactly the classical (i.e. nonquantum) value of the momentum of a particle
with mass $m$ moving along the $z$ axis with the velocity $v$. However, in the opposite case when 
$m\ll \hbar/a$ the transformed state is not semiclassical since the uncertainty of $p_z$ is of the same order
as the mean value of $p_z$.

If the photon mass is exactly zero then the photon cannot have the rest state. However, even if the photon mass
is not exactly zero, it is so small that the condition $m\ll \hbar/a$ is certainly satisfied for any realistic
value of $a$. Hence a semiclassical state for the photon or a particle with a very small mass cannot be obtained
by applying the Lorentz transformation to $\chi_0({\bf p})$ and considering the case when $v$ is very close to unity.
An analogous problem with the covariant description of the massless wave function has been discussed in
the preceding section (see Eq. (\ref{p-p0B})). 

The above discussion shows that in the relativistic case the momentum distribution in transverse directions is the
same as in the nonrelativistic case (see also Eq. (\ref{p-p0B})) and the difference arises only for 
the momentum distribution in the longitudinal direction. Let us consider the ultrarelativistic case when 
$|{\bf p}_0|=p_0\gg m$ and suppose that ${\bf p}_0$ is directed along the $z$ axis. As noted in the preceding section,
the formal requirement of Lorentz covariance will be satisfied if one works with rotationally invariant 
combinations of ${\bf p}$ and ${\bf p}_0$. The quantities ${\bf p}_{\bot}^2$ and $(p_z-p_0)^2$ satisfy this condition
because
$${\bf p}_{\bot}^2=[{\bf p}-{\bf p}_0\frac{({\bf p}{\bf p}_0)}{p_0^2}]^2, \quad (p_z-p_0)^2=\frac{1}{p_0^2}[({\bf p}{\bf p}_0)-p_0^2]^2$$

We will describe an ultrarelativistic semiclassical state by a wave function 
which is a generalization of the function (\ref{chip}) (see also Eq. (\ref{covxi})):
\begin{equation}
\chi({\bf p},0)=\frac{ab^{1/2}}{\pi^{3/4}\hbar^{3/2}}exp[-\frac{{\bf p}_{\bot}^2a^2}{2\hbar^2}
-\frac{(p_z-p_0)^2b^2}{2\hbar^2}-\frac{i}{\hbar}{\bf p}_{\bot}{\bf r}_{0\bot}-\frac{i}{\hbar}(p_z-p_0)z_0]
\label{chiprel}
\end{equation}
In the general case the parameters $a$ and $b$ defining the momentum distributions in the transverse and 
longitudinal  directions, respectively, can be different. In that case the uncertainty of each transverse component of momentum is 
$\hbar /(a\sqrt{2})$ while the uncertainty of the $z$ component of momentum is $\hbar /(b\sqrt{2})$.
In view of the above discussion one might think that, as a consequence of the Lorentz contraction, the parameter $b$
should be very small. However, the notion of the Lorentz contraction has a physical
meaning only if $m\gg \hbar/a$ while for the photon the opposite relation takes place. We will see below
that in typical situations the quantity $b$ is large and much greater than $a$.

In relativistic quantum theory the situation with time is analogous to that in the nonrelativistic case (see Sec.
\ref{NRWPS}) and time can be treated only as a good approximate parameter describing the evolution according to the Schr\"{o}dinger equation with the
relativistic Hamiltonian. Then, as a consequence of Eq. (\ref{chiptA}), we have that in the ultrarelativistic
case (i.e. when $p=|{\bf p}|\gg m$) 
\begin{equation}
\chi({\bf p}, t)=exp(-\frac{i}{\hbar}pct)\chi({\bf p},0)
\label{chiptphoton}
\end{equation}
Since at different moments
of time the wave functions in momentum space differ each other only by a phase factor, the mean value and uncertainty
of each momentum component do not depend on time. In other words, there is no WPS for the wave function in momentum
space. As noted in Sec. \ref{NRWPS}, the same is true in the nonrelativistic case.

As noted in the preceding section, in the relativistic case the function $\psi({\bf r},t)$ can be again defined by Eq. (\ref{Fourier})
where now $\chi({\bf p}, t)$ is defined by Eq. (\ref{chiptphoton}). If the variable $p_z$ in the 
integrand is replaced by $p_0+p_z$ then as follows from Eqs. (\ref{Fourier},\ref{chiprel},\ref{chiptphoton})
\begin{eqnarray}
&&\psi({\bf r},t)=\frac{ab^{1/2}exp(i{\bf p}_0{\bf r}/\hbar)}{\pi^{3/4}\hbar^{3/2}(2\pi\hbar)^{3/2}}
\int exp\{-\frac{{\bf p}_{\bot}^2a^2}{2\hbar^2}-\frac{p_z^2b^2}{2\hbar^2}+\frac{i}{\hbar}{\bf p}({\bf r}-{\bf r}_0)\nonumber\\
&&-\frac{ict}{\hbar}[(p_z+p_0)^2+{\bf p}_{\bot}^2]^{1/2}\} d^3{\bf p}
\label{psirtphoton}
\end{eqnarray}

In contrast to the nonrelativistic case where the energy is the quadratic function of momenta and the integration in
 Eq. (\ref{psirt}) can be performed analytically, here the analytical integration is a problem in
view of the presence of square root in Eq. (\ref{psirtphoton}). We will perform the integration
by analogy with the Fresnel approximation in optics and with Ref. \cite{Dillon} where a
similar approximation has been used for discussing the WPS effect in classical electrodynamics.
The Fresnel approximation describes some important features of the relativistic WPS
effect but, as will be noted below, in this approximation some important features of this effect
are lost.   

The approximation is based on the fact that in semiclassical approximation the quantity $p_0$ should be much greater
than uncertainties of the momentum in the longitudinal and transversal directions, i.e. $p_0\gg p_z$ and
$p_0\gg |{\bf p}_{\bot}|$. Hence with a good accuracy one can expand the square 
root in the integrand in
powers of $|{\bf p}|/p_0$. Taking into account the linear and quadratic terms in the square root we get
\begin{equation}
[(p_z+p_0)^2+{\bf p}_{\bot}^2]^{1/2}\approx p_0+p_z+{\bf p}_{\bot}^2/2p_0
\label{pperp}
\end{equation}
This is analogous to the approximation 
$(m^2+{\bf p}^2)^{1/2}\approx m+{\bf p}^2/2m$ in nonrelativistic case.
Then the integral over $d^3{\bf p}$ can be calculated as a product of integrals over $d^2{\bf p}_{\bot}$
and $dp_z$ and the calculation is analogous to that in Eq. (\ref{psirt}). The result of the calculation is
\begin{eqnarray}
&&\psi({\bf r},t)=[\pi^{3/4}ab^{1/2}(1+\frac{i\hbar ct}{p_0a^2})]^{-1}
exp[\frac{i}{\hbar}({\bf p}_0{\bf r}-p_0ct)]\nonumber\\
&&exp[-\frac{ ({\bf r}_{\bot}-{\bf r}_{0\bot})^2(1-\frac{i\hbar ct}{p_0a^2})}{2a^2(1+\frac{\hbar^2c^2t^2}{p_0^2a^4})}
-\frac{(z-z_0-ct)^2}{2b^2}]
\label{final}
\end{eqnarray}

This result shows that the wave packet describing an ultrarelativistic particle (including a photon) is moving
along the classical trajectory $z(t)=z_0+ct$, in the longitudinal direction there is no spreading while in
transverse directions spreading is characterized by the function 
\begin{equation}
a(t)=a(1+\frac{\hbar^2c^2t^2}{p_0^2a^4})^{1/2}
\label{at}
\end{equation}
The characteristic time of spreading can be defined as $t_*=p_0a^2/\hbar c$. 
The fact that $t_*\to \infty$ in the
formal limit $\hbar\to 0$ shows that in relativistic case WPS also is a pure quantum phenomenon (see the end of
Sec. \ref{NRWPS}). From the formal point of view the result for $t_*$ is the same as in nonrelativistic
theory but $m$ should be replaced by $E/c^2$ where $E$ is the energy of the ultrarelativistic particle. 
This fact could be expected since, as noted above, it is reasonable to think that spreading in directions perpendicular 
to the particle momentum is similar to that in standard nonrelativistic
quantum mechanics. However,
in the ultrarelativistic case spreading takes place only in these directions.
If $t\gg t_*$ the transverse width
of the packet is $a(t)=\hbar ct/(p_0a)$. 

Hence the speed of spreading in perpendicular directions is $v_*=\hbar c/p_0a$.
In the nonrelativistic case different points of the packet are moving with different
velocities and this is not a problem but in the case of the photon one expects
that each point is moving with the speed $c$. However, the Fresnel approximation
creates a problem because different points are moving with different velocities
such that their magnitudes are in the range $[c,(c^2+v_*^2)^{1/2}]$. 

We now consider a model where 
\begin{equation}
\chi({\bf p})=f({\bf p}/p)F(p)/p
\label{relchipn}
\end{equation}
and assume that $f({\bf p}/p)=\sum_{l\mu}c_{l\mu}Y_{l\mu}({\bf p}/p)$ is the
decomposition of the function $f$ over spherical functions.  The dependence
of the momentum wave function on $t$ is now defined by Eq. (\ref{chiptphoton}).
In full analogy with the derivation of Eq. (\ref{psirn}) we now get that
\begin{eqnarray}
&&\psi({\bf r},t)=\frac{-i}{(2\pi\hbar)^{1/2}r}\sum_{l\mu}c_{l\mu}Y_{l\mu}({\bf r}/r)
[G(ct-r)-(-1)^lG(ct+r)]
\label{relpsirn}
\end{eqnarray}
where
\begin{equation}
G(\xi)=\int_0^{\infty} F(p)exp(\frac{-i}{\hbar}\xi p )dp
\label{Gxi}
\end{equation}

For reasonable choices of $F(p)$ we will have that at large distances and times $G(ct-r)\gg G(ct+r)$. 
Indeed if, for example, the quantities $p_0$ and $b$ are such that $p_0b\gg \hbar$ then
possible $(F,G)$ choices are: 
\begin{eqnarray}
&&F(p)=exp(-\frac{|p-p_0|b}{\hbar}),\quad G(\xi)=\frac{exp(-ip_0\xi/\hbar)}{b^2+\xi^2};\nonumber\\
&&F(p)=exp(-\frac{(|p-p_0|b)^2}{2\hbar^2}),\quad 
G(\xi)=(2\pi)^{1/2}\frac{\hbar}{b}exp(-\frac{ip_0\xi}{\hbar}-\frac{\xi^2}{2b^2})
\label{FG}
\end{eqnarray}
As follows from Eq. (\ref{relpsirn}), in those cases
\begin{eqnarray}
\psi({\bf r},t)=\frac{-i}{(2\pi\hbar)^{1/2}r}f({\bf r}/r)G(ct-r)
\label{relpsif}
\end{eqnarray}
Therefore at each moment of time $t$ the coordinate wave function is not negligible only
inside a narrow sphere with the radius $ct$ and the width of the order of $b$.

The conclusion is that, in contrast to the nonrelativistic case, in the ultrarelativistic one 
there is no WPS in the radial direction (by analogy with the Fresnel approximation) and,
by analogy with the result (\ref{psif}), at large distances and times the angular distributions
in momentum and coordinate wave functions are the same. Therefore, in full analogy with
the Mott-Heisenberg problem (see Sec. \ref{Mott}), the momenta
of particles detected by a measuring device will be in the angular range defined not by
the function $f({\bf r}/r)$ but by the function ${\tilde f}({\bf r}/r)$ characterizing the
angles at which the device is seen from the origin. In addition, the angular distribution of
momenta characterized by the function $f$ does not depend on time, as well as in the
nonrelativistic case.

If the function $f$ is essentially different from zero only in the range where 
angles between momenta and the $z$-axis are small then the model (\ref{relchipn}) gives
the same qualitative predictions as the Fresnel approximation. Indeed, suppose that
this function is essentially different from zero for angles which are of the order of
$\alpha$ or less, and $\alpha\ll 1$. Then the parameter $b$ in Eq. (\ref{FG}) is
similar to the parameter $b$  in Eq. (\ref{chiprel}). The characteristic magnitude of the transverse
momentum is of the order of $p_{\bot}\approx \alpha p_0$. Let $a$ be defined such that 
$p_{\bot}=\hbar/a$. When the time is greater than a characteristic time
for which the transition from Eq. (\ref{relpsirn}) to Eq. (\ref{relpsif}) is legitimate
(this time can differ from $t_*$ for the Fresnel model) then, since the angular
distributions in the momentum and coordinate wave functions are the same, the
transversal width of the packet is of the order of $\alpha ct\approx ct\hbar/(p_0a)$ in
agreement with the Fresnel approximation. Therefore {\it if $t$ is greater than some
characteristic time then the width $a(t)$ of the packet is inversely proportional to
the initial width $a(0)=a$.} It is also possible to define $v_*$ by the same expression as
in the Fresnel approximation. If $v_*\ll c$ the only difference between the two models
is that in the Fresnel approximation different points of the packet are moving with different
speeds while in the model (\ref{relchipn}) they are moving with the same speed $c$.
In fact the Fresnel approximation is such that a small arc representing the front of the wave function
 in the model (\ref{relchipn}) is replaced by a segment.

\section{Geometrical optics}
\label{geom}

The relation between quantum and classical electrodynamics is well-known 
and is described in textbooks (see e.g. Ref. \cite{Dirac,AB}). As already 
noted, classical electromagnetic field consists of many photons and in 
classical electrodynamics the photons are not described individually. 
Instead, classical electromagnetic field is described by field strengths
which represent mean characteristics of a large set of photons. For 
constructing the field strengths one can use the photon wave functions 
$\chi({\bf p},t)$ or
$\psi({\bf r},t)$ where $E$ is replaced by $\hbar\omega$ and ${\bf p}$ 
is replaced by $\hbar {\bf k}$. In this connection it is interesting to note that
since $\omega$ is a classical quantity used for describing a classical electromagnetic field, 
the photon is a pure quantum
particle since its energy disappears in the formal limit $\hbar\to 0$. Even this
fact shows that the photon cannot be treated as a classical particle and the effect of WPS 
for the photon cannot be neglected.

With the above replacements the functions $\chi$ and $\psi$ do not contain any 
dependence on $\hbar$ (note that the normalization factor $\hbar^{-3/2}$ 
in $\chi({\bf k},t)$ will disappear since the normalization integral for 
$\chi({\bf k},t)$ is now over $d^3{\bf k}$, not $d^3{\bf p}$). The 
quantities $\omega$ and ${\bf k}$ are now treated, respectively, as the 
frequency and the wave vector of the classical electromagnetic field, and 
the functions $\chi({\bf k},t)$ and $\psi({\bf r},t)$ are interpreted 
not such that they describe probabilities for a single photon but such that
they describe classical electromagnetic field ${\bf E}({\bf r},t)$ 
and ${\bf B}({\bf r},t)$ which be constructed from these functions as
described in textbooks on QED (see e.g. Ref. \cite{AB}).

An additional argument in favor of the choice of $\psi({\bf r},t)$ as the coordinate photon wave function
is that in classical electrodynamics the 
quantities ${\bf E}({\bf r},t)$ and ${\bf B}({\bf r},t)$ for the free 
field should satisfy the wave equation $\partial^2{\bf E}/c^2 \partial t^2=\Delta {\bf E}$ and 
analogously for ${\bf B}({\bf r},t)$. Hence if 
${\bf E}({\bf r},t)$ and ${\bf B}({\bf r},t)$ are constructed from $\psi({\bf r},t)$ as described
in textbooks (see e.g. Ref. \cite{AB}), they will satisfy the wave equation since, as follows from
Eqs. (\ref{Fourier},\ref{chiprel},\ref{chiptphoton}), $\psi({\bf r},t)$ also satisfies this equation.

The geometrical optics approximation
implies that if ${\bf k}_0$ and ${\bf r}_0$ are the mean values
of the wave vector and the spatial radius vector for a wave packet describing the electromagnetic wave then
the uncertainties $\Delta k$ and $\Delta r$, which are the mean values of $|{\bf k}-{\bf k}_0|$ and
$|{\bf r}-{\bf r}_0|$, respectively, should satisfy the requirements $\Delta k\ll |{\bf k}_0|$ and
$\Delta r\ll |{\bf r}_0|$. In full analogy with the derivation of Eq. (\ref{uncert}), one
can show that for each $j=1,2,3$ the uncertainties of the corresponding projections of the vectors
${\bf k}$ and ${\bf r}$ satisfy the requirement $\Delta k_j\Delta r_j\geq 1/2$
(see e.g. Ref. \cite{LLII}). In particular, an electromagnetic wave satisfies the approximation of geometrical optics
in the greatest possible extent if $\Delta k\Delta r$ is of the order of unity.

The above discussion confirms what has been mentioned in Sec. \ref{intropos} that {\it the effect of WPS in transverse directions takes place not only in quantum theory
but even in classical electrodynamics}. Indeed, since the function $\psi({\bf r},t)$ satisfies the classical wave equation,
the above consideration can be also treated as an example showing that {\it even for a free wave packet in classical
electrodynamics the WPS effect is inevitable}. In the language of classical waves the parameters of spreading can be
characterized by the function $a(t)$ (see Eq. (\ref{at})) and the quantities $t_*$ and $v_*$ such that in terms of the wave length $\lambda=2\pi c/\omega_0$ 
\begin{equation}
a(t)=a(1+\frac{\lambda^2c^2t^2}{4\pi^2a^4})^{1/2},\quad t_*=\frac{2\pi a^2}{\lambda c}, \quad 
v_*=\frac{\lambda c}{2\pi a}
\label{em}
\end{equation}
The last expression can be treated such that if $\lambda\ll a$ then the momentum has the angular
uncertainty of the order of $\alpha=\lambda/(2\pi a)$. This result is natural from the following consideration.
Let the mean value of the momentum be directed along the $z$-axis and the uncertainty of the transverse 
component of the
momentum be $\Delta p_{\bot}$. Then $\Delta p_{\bot}$ is of the order of $\hbar/a$, $\lambda=2\pi\hbar/p_0$ and 
hence $\alpha$ is of the order of 
$\Delta p_{\bot}/p_0\approx \lambda/(2\pi a)$. This is analogous to the well-known result in classical optics
that the best angular resolution of a telescope with the dimension $d$ is of the order of $\lambda/d$.
Another well-known result of classical optics is that if a wave encounters an obstacle having the dimension $d$
then the direction of the wave diverges by the angle of the order of $\lambda/d$. 

The inevitability of WPS for a free wave packet in classical electrodynamics is obvious from the following
consideration. Suppose that a classical wave packet does not have a definite value of the momentum. 
Then if $a$ is the initial width of the packet
in directions perpendicular to the mean momentum, one might expect that the width will grow as $a(t)=a+\alpha ct$
and for large values of $t$, $a(t)\approx \alpha ct$. As follows from Eq. (\ref{em}), if $t\gg t_*$ then indeed
$a(t)\approx \alpha ct$. In standard quantum theory we have the same result because the coordinate and momentum wave 
functions are related to each other by the same Fourier transform as the coordinate and ${\bf k}$ distributions
in classical electrodynamics. 

The quantity $N_{||}=b/\lambda$ shows how many oscillations the oscillating exponent in Eq. (\ref{final})
makes in the region where the wave function or the amplitude of the classical wave is significantly different
from zero. As noted in Sec. \ref{classical}, for the validity of semiclassical approximation this quantity should be 
very large. In nonrelativistic quantum mechanics $a$ and $b$ are of the same order and hence the same can be said
about the quantity $N_{\bot}=a/\lambda$. As noted above, in the case of the photon we do not know the relation between $a$
and $b$. In terms of the quantity $N_{\bot}$ we can rewrite the expressions for $t_*$ and $v_*$ in Eq. (\ref{em}) as
\begin{equation}
t_*=2\pi N_{\bot}^2 T,\quad v_*=\frac{c}{2\pi N_{\bot}}
\label{tv}
\end{equation}
where $T$ is the period of the classical wave. Hence the accuracy of semiclassical approximation 
(or the geometrical optics approximation in classical
electrodynamics) increases with the increase of $N_{\bot}$.

In Ref. \cite{Dillon} the problem of WPS for classical electromagnetic waves has been discussed 
in the Fresnel approximation for a two-dimensional wave
packet. Equation (25) of Ref. \cite{Dillon} is a special case of Eq. (\ref{pperp}) and the author of 
Ref. \cite{Dillon} shows
that, in his model the wave packet spreads out in the direction perpendicular to the group velocity of the
packet. As noted in the preceding section,
in the ultrarelativistic case the function $a(t)$ is given by the same expression as in the nonrelativistic case
but $m$ is replaced by $E/c^2$. Hence if the results of the preceding section are reformulated in terms of
classical waves then $m$ should be replaced by $\hbar\omega_0/c^2$ and this fact has been pointed out in Ref. \cite{Dillon}.

\section{Wave packet width paradox}
\label{WPW}

We now consider the following important question. We assume that a classical wave packet is a collection of photons.
Let $a_{cl}$ be the quantity $a$ for the classical packet and $a_{ph}$ be a typical value of $a$ for the photons.
What is the relation between $a_{cl}$ and $a_{ph}$? 

My observation is that physicists answer this question in different ways. Quantum physicists usually say  
that in typical situations $a_{ph}\ll a_{cl}$ because $a_{cl}$ is of macroscopic size while in semiclassical
approximation the quantity $a_{ph}$ for each photon can be treated as the size of the region where the photon has 
been created. On the other hand, classical physicists usually say that
$a_{ph}\gg a_{cl}$ and the motivation follows.

Consider a decomposition of some component of classical electromagnetic field into the Fourier series: 
\begin{equation}
A(x)=\sum_{\sigma} \int [a({\bf p},\sigma)u({\bf p},\sigma)exp(-ipx)+
a({\bf p},\sigma)^*u({\bf p},\sigma)^*exp(ipx)]d^3{\bf p}
\label{Acont}
\end{equation}
where $\sigma$ is the polarization, $x$ and $p$ are the four-vectors such that $x=(ct,{\bf x})$ and $p=(|{\bf p}|c,{\bf p})$,
the functions $a({\bf p},\sigma)$ are the same for all the components, the functions $u({\bf p},\sigma)$
depend on the component and $^*$ is used to denote the complex conjugation. Then photons arise as a result of quantization 
when $a({\bf p},\sigma)$ and $a({\bf p},\sigma)^*$ are understood
not as usual function but as operators of annihilation and creation of the photon with the quantum numbers
$({\bf p},\sigma)$ and $^*$ is now understood as Hermitian conjugation. Hence the photon is described by a plane wave 
which has the same magnitude in all points of the
space. In other words, $a_{ph}$ is infinitely large and a finite width of the classical wave packet arises as a result
of interference of different plane waves. 

The above definition of the photon has at least the following inconsistency. If the photon is treated as a particle
then its wave function should be normalizable while the plane wave is not normalizable. In textbooks this problem
is often circumvented by saying that we consider our system in a finite box. Then the spectrum of momenta becomes
finite and instead of Eq. (\ref{Acont}) one can write
\begin{equation}
A(x)=\sum_{\sigma} \sum_j [a({\bf p}_j,\sigma)u({\bf p}_j,\sigma)exp(-ip_jx)+a({\bf p}_j,\sigma)^*
u({\bf p}_j,\sigma)^*exp(ip_jx)]
\label{Adiscr}
\end{equation}
where $j$ enumerates the points of the momentum spectrum.

One can now describe quantum electromagnetic field by states in the Fock space where the vacuum vector $\Phi_0$ 
satisfies the condition $a({\bf p}_j,\sigma)\Phi_0=0$, $||\Phi_0||=1$ and the operators commute as
\begin{equation}
[a({\bf p}_i,\sigma_k),a({\bf p}_j,\sigma_l)]=[a({\bf p}_i,\sigma_k)^*,a({\bf p}_j,\sigma_l)^*]=0,
\quad [a({\bf p}_i,\sigma_k),a({\bf p}_j,\sigma_l)^*]=\delta_{ij}\delta_{kl}
\label{adiscr}
\end{equation}
Then any state can be written as
\begin{equation}
\Psi=\sum_{n=0}^{\infty}\sum_{\sigma_1...\sigma_n}\sum_{{\bf p}_1,...{\bf p}_n} 
\chi({\bf p}_1,\sigma_1,...{\bf p}_n,\sigma_n)
a({\bf p}_1,\sigma_1)^*\cdots a({\bf p}_n,\sigma_n)^* \Phi_0
\label{Psidiscrt}
\end{equation}

Classical states are understood such that although the number of photons is large, it is much less than the number
of possible momenta and in Eq. (\ref{Psidiscrt}) all the photons have different momenta (this is analogous to
the situation in classical statistics where mean occupation numbers are much less than unity).
Then it is not important whether the operators $(a,a^*)$ commute or anticommute. However, according to the Pauli theorem on
spin-statistics connection \cite{Pauli2},
they should commute and this allows the existence of coherent states where many photons 
have the same quantum numbers. Such states can be created in lasers and they are not described by classical
electrodynamics. In the next section we consider position operator for coherent states while in this section we
consider only quantum description of states close to classical.

Note that even in some textbooks on quantum optics (see e.g. Ref. \cite{Mandel}) classical and quantum states
are characterized in the opposite way: it is stated that classical states are characterized by large occupation numbers
while quantum states - by small ones. The question what states should be called classical or quantum is not a matter of
convention since in quantum theory there are rigorous criteria for that purpose. In particular, as explained
in textbooks on quantum theory, the exchange interaction is a pure quantum phenomenon which does not have classical
analogs. That's why the Boltzmann statistics (which works when mean occupation numbers are much less than unity
and the exchange interaction is negligible) is classical while the Fermi-Dirac and Bose-Einstein statistics (which
work when mean occupation numbers are of the order of unity or greater and the exchange interaction is important)
are quantum.  

The next problem is that one should take into account that in standard theory the photon momentum spectrum is
continuous. Then the above construction can be generalized as follows. The vacuum state $\Phi_0$ satisfies the
same conditions $||\Phi_0||=1$ and $a({\bf p},\sigma)\Phi_0=0$ while the operators $(a,a^*)$ satisfy the following
commutation relations
\begin{equation}
[a({\bf p},\sigma),a({\bf p}',\sigma')]=[a({\bf p},\sigma)^*,a({\bf p}',\sigma')^*]=0,\quad  
[a({\bf p},\sigma),a({\bf p}',\sigma')^*]=\delta^{(3)}({\bf p}-{\bf p}')\delta_{\sigma\sigma'}
\label{photonaa*}
\end{equation}
Then a general quantum state can be written as
\begin{equation}
\Psi=\sum_{n=0}^{\infty}\sum_{\sigma_1...\sigma_n}\int ...\int \chi({\bf p}_1,\sigma_1,...{\bf p}_n,\sigma_n)
a({\bf p}_1,\sigma_1)^*\cdots a({\bf p}_n,\sigma_n)^* d^3{\bf p}_1\cdots d^3{\bf p}_n \Phi_0
\label{manyphoton}
\end{equation}

In the approximation when a classical wave packet is understood as a collection of independent photons
(see the discussion in Sec. \ref{Discussion}), the state of this packet has the form 
\begin{equation}
\Psi=\sum_{n=0}^{\infty}c_n \prod_{j=1}^n \{\sum_{\sigma_j}\int\chi_j({\bf p}_j,\sigma_j)
a({\bf p}_j,\sigma_j)^*d^3{\bf p}_j\}
\Phi_0
\label{indepphotons}
\end{equation}
where $\chi_j$ is the wave function of the $j$th photon and intersections of supports of wave functions of different photons
can be neglected. This is an analog of the above situation with the discrete case where it is assumed that 
different photons in a classical wave packet have different momenta. In other words, while the wave function of
each photon can be treated as an interference of plane waves, different photons can interfere only
in coherent states but not in classical wave packets. 

We now describe a well-known generalization of the results on IRs of the Poincare algebra to the description 
in the Fock space (see e.g. Ref. \cite{JPA} for details). 
If $A$ is an operator in the space of the photon IR then a generalization of this operator to the case of the
Fock space can be constructed as follows. Any operator in the space of IR can be represented as an integral
operator acting on the wave function as
\begin{equation}
A\chi({\bf p},\sigma)=\sum_{\sigma'}\int A({\bf p},\sigma,{\bf p}',\sigma')\chi({\bf p}',\sigma')d^3{\bf p}'
\label{integraloper}
\end{equation} 
For example, if ${\bf A}\chi({\bf p},\sigma)=\partial \chi({\bf p},\sigma)/\partial{\bf p}$ then ${\bf A}$ is
the integral operator with the kernel 
$${\bf A}({\bf p},\sigma,{\bf p}',\sigma')=\frac{\partial 
\delta^{(3)}({\bf p}-{\bf p}')}{\partial{\bf p}}\delta_{\sigma\sigma'}$$
We now require that if the action of the operator $A$ in the space of IR is defined by Eq. (\ref{integraloper})
then in the case of the Fock space this action is defined as
\begin{equation}
A=\sum_{\sigma\sigma'}\int A({\bf p},\sigma,{\bf p}',\sigma')a({\bf p},\sigma)^*a({\bf p}',\sigma')
d^3{\bf p}d^3{\bf p}'
\label{Fockoper}
\end{equation}
Then it is easy to verify that if $A$, $B$ and $C$ are operators in the space of IR satisfying the commutation
relation $[A,B]=C$ then the generalizations of these operators in the Fock space satisfy the same commutation relation.
It is also easy to verify that the operators generalized to the action in the Fock space in such a way are additive,
i.e. for a system of $n$ photons they are sums of the corresponding single-particle operators. In particular, the
energy of the $n$-photon system is a sum of the energies of the photons in the system and analogously for the other
representation operators of the Poincare algebra - momenta, angular momenta and Lorentz boosts.

We are interested in calculating mean values of different combinations of the momentum operator. Since this operator
does not act over spin variables, we will drop such variables in the $(a,a^*)$ operators and in the functions $\chi_j$.
Then the explicit form of the momentum operator is 
${\bf P}=\int {\bf p} a({\bf p})^*a({\bf p})d^3{\bf p}$. Since this operator does not change the number of photons,
the mean values can be independently calculated in each subspace where the number of photons is $N$.

Suppose that the momentum of each photon is
approximately directed along the $z$-axis and the quantity $p_0$ for each photon approximately equals
$2\pi\hbar/\lambda$. If $\Delta p_{\bot}$ is a typical uncertainty of the transversal component of the
momentum for the photons then a typical value of the angular uncertainty for the photons is 
$\alpha_{ph}=\Delta p_{\bot}/p_0\approx \lambda/(2\pi a_{ph})$. The total momentum of the classical wave packet
consisting of $N$ photons is a sum of the photon momenta: ${\bf P}=\sum_{i=1}^N {\bf p}^{(i)}$. 
Suppose that the mean value of ${\bf P}$
is directed along the $z$-axis and its magnitude $P_0$ is such that $P_0\approx Np_0$. The uncertainty of the
$x$ component of ${\bf P}$ is $\Delta P_x={\overline{P_x^2}}^{1/2}$ where
$${\overline{P_x^2}}=\sum_{i=1}^N \overline{(p_x^{(i)})^2}+\sum_{i\neq j;i,j=1}^N\overline{p_x^{(i)}p_x^{(j)}}$$
Then in the approximation of independent photons (see the remarks after Eq. (\ref{indepphotons}))
$${\overline{P_x^2}}=\sum_{i=1}^N \overline{(p_x^{(i)})^2}+\sum_{i\neq j;i,j=1}^N\overline{p_x^{(i)}}\cdot
\overline{p_x^{(j)}}=\sum_{i=1}^N [\overline{(p_x^{(i)})^2}-\overline{p_x^{(i)}}^2]=
\sum_{i=1}^N (\Delta p_x^{(i)})^2$$
where we have taken into account that $\overline{P_x}=\sum_{i=1}^N\overline{p_x^{(i)}}=0$.

As a consequence, if typical values of $\Delta p_{\bot}^{(i)}$ have the the same order of magnitude equal 
to $\Delta p_{\bot}$
then $\Delta P_{\bot}\approx N^{1/2} \Delta p_{\bot}$ and the angular divergence of 
the classical vave packet is 
\begin{equation}
\alpha_{cl}=\Delta P_{\bot}/P_0\approx \Delta p_{\bot}/(p_0N^{1/2})=\alpha_{ph}/N^{1/2}
\label{acl}
\end{equation}
Since the classical wave packet is described by the same wave equation as the photon wave function, its angular
divergence can be expressed in terms of the parameters $\lambda$ and $a_{cl}$ such that 
$\alpha_{cl}=\lambda/(2\pi a_{cl})$. Hence $a_{cl}\approx N^{1/2}a_{ph}$ and we conclude that $a_{ph}\ll a_{cl}$.

Note that in this derivation no position operator has been used. Although the quantities $\lambda$ and $a_{ph}$ have
the dimension of length, they are defined only from considering the photon in momentum space because, as noted in Sec.
\ref{momentum}, for individual photons $\lambda$ is understood only as $2\pi\hbar/p_0$, $a_{ph}$ defines the
width of the photon momentum wave function (see Eq. (\ref{chiprel})) and is of the order of $\hbar/\Delta p_{\bot}$.
As noted in Secs. \ref{NRWPS} and \ref{RelWPS}, the momentum distribution does not depend on time and hence the
result $a_{ph}\ll a_{cl}$ does not depend on time too. If photons in a classical wave packet could be treated as 
(almost) pointlike particles then photons do not experience WPS while the WPS effect for a classical wave packet 
is a consequence of the fact that different photons in the packet have different momenta. 

However, in standard quantum theory this scenario does not take place for the following reason. 
Let $a_{cl}(t)$ be the quantity $a(t)$ for the classical wave packet and $a_{ph}(t)$ be a typical value of the
quantity $a(t)$ for individual photons. With standard position operator the quantity $a_{ph}(t)$ is interpreted
as the spatial width of the photon coordinate wave function in directions perpendicular to the photon
momentum and this quantity is time dependent. As shown in Secs. \ref{RelWPS} and \ref{geom}, $a(0)=a$ but if 
$t\gg t_*$ then $a(t)$ is {\it inversely proportional} to $a$ and the coefficient of
proportionality is the same for the classical wave packet and individual photons (see Eq. (\ref{em})). 
Hence {\it in standard quantum theory we have a paradox that after some period of time} $a_{ph}(t)\gg a_{cl}(t)$
i.e. individual photons in a classical wave packet spread out in a much greater extent than the wave packet
as a whole. We call this situation the wave packet width (WPW) paradox 
(as noted above, different photons in
a classical wave packet do not interfere with each other). The reason of the paradox is obvious:
if the law that the angular divergence of a wave packet is of the order of $\lambda/a$ is applied to both,
a classical wave packet and photons comprising it then the paradox follows from the fact that the quantities $a$ for
the photons are much less than the quantity $a$ for the classical wave packet. Note that in classical case the 
quantity $a_{cl}$ does not have the meaning of $\hbar/\Delta P_{\bot}$ and $\lambda$ is not equal to $2\pi\hbar/P_0$.

\section{Wave packet spreading in coherent states}
\label{coherent}

In textbooks on quantum optics the laser emission is described by the following model (see e.g. 
Refs. \cite{Mandel,Scully}). 
Consider a set of photons having the same momentum ${\bf p}$ and polarization $\sigma$ and, by analogy with the discussion
in the preceding section, suppose that the momentum spectrum is discrete. 
Consider a quantum superposition $\Psi=\sum_{n=0}^{\infty} c_n [a({\bf p},\sigma)^*]^n\Phi_0$ where the coefficients $c_n$ satisfy 
the condition that $\Psi$ is an eigenstate of the annihilation operator $a({\bf p},\sigma)$. Then 
the product of the coordinate and momentum uncertainties has the minimum possible value $\hbar/2$ and, as 
noted in Sec. \ref{classical}, such a state is called coherent. However, the term coherent is sometimes used meaning
that the state is a quantum superposition of many-photon states $[a({\bf p},\sigma)^*]^n\Phi_0$.

In the above model it is not taken into account that (in standard theory) photons emitted by a laser can have only a
continuous spectrum of momenta. Meanwhile for the WPS effect the width of the momentum
distribution is important. In this section we consider a generalization of the above model where the fact that photons 
have a continuous spectrum of momenta is taken into account. This will make it possible to consider the WPS effect
in coherent states.

In the above formalism coherent states can be defined as follows. We assume that all the photons in the state  
Eq. (\ref{manyphoton}) have the same polarization. Hence for describing such states we can drop the quantum number
$\sigma$ in wave functions and $a$-operators. We also assume that all photons in coherent states have the same momentum
distribution. These conditions can be satisfied by requiring that coherent states have the form 
\begin{equation}
\Psi=\sum_{n=0}^{\infty}c_n[\int\chi({\bf p})a({\bf p})^*d^3{\bf p}]^n \Phi_0
\label{cohstate}
\end{equation} 
where $c_n$ are some coefficients. Finally, by analogy with the description of coherent states in standard textbooks
on quantum optics one can require that they are eigenstates of the operator $\int a({\bf p})d^3{\bf p}$. 

The dependence of the state $\Psi$ in Eq. (\ref{cohstate}) on $t$ is $\Psi(t)=exp(-iEt/\hbar)\Psi$ where, as follows
from Eqs. (\ref{IRoperators}) and (\ref{Fockoper}), the action of the energy operator in the Fock space is
$E=\int pc a({\bf p})^*a({\bf p})d^3{\bf p}$. Since $exp(iEt/\hbar)\Phi_0=\Phi_0$, it readily follows from
Eq. (\ref{photonaa*}) that 
\begin{equation}
\Psi(t)=\sum_{n=0}^{\infty}c_n[\int\chi({\bf p},t)a({\bf p})^*d^3{\bf p}]^n \Phi_0
\label{cohstatet}
\end{equation}
where the relation between $\chi({\bf p},t)$ and $\chi({\bf p})=\chi({\bf p},0)$ is given by Eq. (\ref{chiptphoton}). 

A problem arises how to define the position operator in the Fock space. If this operator is defined by analogy with
the above construction then we get an unphysical result that each coordinate of the $n$-photon system as a whole is
a sum of the corresponding coordinates of the photons in the system. This is an additional argument that the position
operator is less fundamental than the representation operators of the Poincare algebra and its action should be 
defined from additional considerations. In textbooks on quantum optics the position operator for coherent states 
is usually defined by analogy with the position operator in nonrelativistic quantum mechanics for the harmonic
oscillator problem. The motivation follows. If the energy levels $\hbar\omega (n+1/2)$ of the harmonic oscillator
are treated as states of $n$ quanta with the energies $\hbar\omega$ then the harmonic oscillator problem can be described
by the operators $a$ and $a^*$ which are expressed in terms of the one-dimensional position and momentum operators
$q$ and $p$ as $a=(\omega q+ip)/(2\hbar\omega)^{1/2}$ and $a^*=(\omega q-ip)/(2\hbar\omega)^{1/2}$, respectively. However, as noted
above, the model description of coherent states in those textbooks is one-dimensional because the continuous nature
of the momentum spectrum is not taken into account. In addition, the above results on WPS give indications that 
the position operator in standard theory is not consistently defined. For all these reasons a problem arises whether
the requirement that the state $\Psi$ in Eq. (\ref{cohstate}) is an eigenvector of the operator $\int a({\bf p})d^3{\bf p}$
has a physical meaning. In what follows this requirement is not used.

In nonrelativistic classical mechanics the radius vector of a system of $n$ particles as a whole (the radius vector of
the center of mass) is defined as ${\bf R}=(m_1{\bf r}_1+...+m_n{\bf r}_n)/(m_1+...+m_n)$ and in works on 
relativistic classical mechanics it is usually defined as ${\bf R}=(\epsilon_1({\bf p}_1){\bf r}_1+...
+\epsilon_n({\bf p}_n){\bf r}_n)/(\epsilon_1({\bf p}_1)+...+\epsilon_n({\bf p}_n))$ where 
$\epsilon_i({\bf p}_i)=(m_i^2+{\bf p}_i^2)^{1/2}$. Hence if all the particles have the same masses and momenta,
${\bf R}=({\bf r}_1+...+{\bf r}_n)/n$.

These remarks make it reasonable to define the position operator for coherent states as follows. Let $x_j$ be the
$j$th component of the position operator in the space of IR and $A_j({\bf p},{\bf p}')$ be the kernel of this
operator. Then in view of Eq. (\ref{Fockoper}) the action of the operator $X_j$ on the state $\Psi(t)$ in 
Eq. (\ref{cohstate}) can be defined as
\begin{equation}
X_j\Psi(t)=\sum_{n=1}^{\infty}\frac{c_n}{n}\int\int A_j({\bf p}",{\bf p}')a({\bf p}")^*a({\bf p}')
d^3{\bf p}"d^3{\bf p}'[\int\chi({\bf p},t)a({\bf p})^*d^3{\bf p}]^n \Phi_0
\label{Xj}
\end{equation} 

If $\overline{x_j}(t)$ and $\overline{x_j^2}(t)$ are the mean values of the operators $x_j$ and $x_j^2$, 
respectively then as follows from the definition of the kernel of the operator $x_j$
\begin{eqnarray}
&&\overline{x_j}(t)=\int \int \chi({\bf p},t)^*A_j({\bf p},{\bf p}')\chi({\bf p}',t)d^3{\bf p}d^3{\bf p}'\nonumber\\
&&\overline{x_j^2}(t)=\int\int \int \chi({\bf p}",t)^*A_j({\bf p},{\bf p}")^*A_j({\bf p},{\bf p}')
\chi({\bf p}',t)d^3{\bf p}d^3{\bf p}"d^3{\bf p}'
\label{meanxj}
\end{eqnarray} 
and in the case of IR the uncertainty of the quantity $x_j$ is $\Delta x_j(t)=[\overline{x_j^2}(t)-\overline{x_j}(t)^2]^{1/2}$. At the same time, if $\overline{X_j}(t)$ and 
$\overline{X_j^2}(t)$ are the mean values of the operators $X_j$ and $X_j^2$, 
respectively then
\begin{equation}
\overline{X_j}(t)=(\Psi(t),X_j\Psi(t)),\quad \overline{X_j^2}(t)=(\Psi(t),X_j^2\Psi(t))
\label{meanXj}
\end{equation}
and the uncertainty of the quantity $X_j$ is $\Delta X_j(t)=[\overline{X_j^2}(t)-\overline{X_j}(t)^2]^{1/2}$.
Our goal is to express $\Delta X_j(t)$ in terms of $\overline{x_j}(t)$, $\overline{x_j^2}(t)$ and
$\Delta x_j(t)$. 

If the function $\chi({\bf p},t)$ is normalized to one (see Eq. (\ref{norm})) then, as follows from
Eq. (\ref{photonaa*}), $||\Psi(t)||=1$ if
\begin{equation}
\sum_{n=0}^{\infty}n! |c_n|^2=1
\label{normPsi}
\end{equation}
A direct calculation using Eqs. (\ref{photonaa*}), (\ref{Xj}), (\ref{meanxj}) and (\ref{meanXj}) gives
\begin{eqnarray}
&&\overline{X_j}(t)=\overline{x_j}(t)\sum_{n=1}^{\infty}n! |c_n|^2\nonumber\\
&&\overline{X_j^2}(t)=\sum_{n=1}^{\infty}(n-1)!|c_n|^2[\overline{x_j^2}(t)+(n-1)\overline{x_j}(t)^2]
\label{meanxjXj}
\end{eqnarray}
It now follows from Eq. (\ref{normPsi}) and the definitions of the quantities $\Delta x_j(t)$ and 
$\Delta X_j(t)$ that 
\begin{equation}
\Delta X_j(t)^2=(1-|c_0|^2)|c_0|^2\overline{x_j}(t)^2+\sum_{n=1}^{\infty}(n-1)!|c_n|^2\Delta x_j(t)^2
\label{DeltaXj}
\end{equation}
  
Equation (\ref{DeltaXj}) is the key result of this section. It has been derived without using
a specific choice of the single photon position operator. The consequence of this result follows. If the
main contribution to the state $\Psi(t)$ in Eq. (\ref{cohstatet}) is given by
very large values of $n$ then $|c_0|$ is very small and the first term in this expression can be neglected. 
Suppose that the main contribution is given by terms where $n$ is of the order of ${\bar n}$. Then, as follows
from Eqs. (\ref{normPsi}) and (\ref{DeltaXj}), $\Delta X_j(t)$ is of the order of $\Delta x_j(t)/{\bar n}^{1/2}$.
This means that for coherent states where the main contribution is given by very large numbers of photons the effect
of WPS is pronounced in a much less extent than for single photons. 

It is interesting to note that the relation between $\Delta X_j(t)$ and $\Delta x_j(t)$ is analogous to
(\ref{acl}) although those relations describe fully difference situations. In both of them
relative uncertainties for a system of many particles are much less than for a single particle.
Since the WPS effect for photons in laser beams is very small, divergence of the laser beam is
only a consequence of the fact that different photons have different momenta.

\section{Experimental consequences of WPS in standard theory}
\label{experiment}

The problem of explaining the redshift phenomenon has a long history. Different competing approaches can be 
divided into two big sets which we call Theory A and Theory B. In Theory A the redshift has been originally explained 
as a manifestation of the Doppler effect but in recent years  the cosmological and gravitational redshifts 
have been added to the consideration. In this theory the interaction of photons with the interstellar medium is treated as
practically not important. On the contrary, in Theory B, which is often called the tired-light theory, the interaction of photons with the interstellar medium is treated as the main reason for the redshift. At present the majority of physicists 
believe that Theory A 
explains the astronomical data better than Theory B. Even some physicists working on Theory B acknowledged 
that any sort of scattering of light would predict more blurring than is seen (see e.g.
the article "Tired Light" in Wikipedia).

As follows from these remarks, in Theory A it is assumed that with a good accuracy we can treat photons 
as propagating in empty space. It is also reasonable to expect (see the discussion in the next section) 
that photons from
distant stars practically do not interact with each other. Hence the effect of WPS can be considered for 
each photon independently and the results of the preceding sections make it possible to understand what 
experimental consequences of WPS are. 

A question arises what can be said about characteristics of photons coming to Earth from distance objects. 
Since wave lengths of such photons are typically much less than all characteristic dimensions
in question one
might think that the radiation of stars can be described in the geometrical optics approximation. As
discussed in Sec. \ref{geom}, this approximation is similar to semiclassical approximation in quantum
theory. This poses a question whether this radiation can be approximately treated as a collection
of photons moving along classical trajectories. However, as noted below, not all photons in the
radiation can be treated in such a way.

Consider, for example, the Lyman transition $2P\to 1S$ in the hydrogen atom, which plays
an important role in the star radiation. We first consider the case when the atom is at rest.
Then the mean energy of the photon is $E_0=10.2eV$, its wave length is $\lambda=121.6nm$ and
the lifetime is $\tau=1.6\cdot 10^{-9}s$. The phrase that the 
lifetime is $\tau$ is interpreted such that the uncertainty of the energy is
$\hbar/\tau$. This implies that the uncertainty of the momentum magnitude is $\hbar/c\tau$ 
and $b$ is of the order of $c\tau\approx 0.48m$.
In this case the photon has a very narrow energy distribution since the mean value of the momentum
$p_0=E_0/c$ satisfies the condition $p_0b\gg \hbar$. At the same time, since the orbital angular momentum of the photon
is a small quantity, the function $f(\theta)=f({\bf p}/p)$ in Eq. (\ref{FG}) has the same order of magnitude at all angles and the direction of the photon momentum cannot be semiclassical. If the atom is not
at rest those conclusions remain valid because typically the speed of the atom is much less than $c$.

As pointed out in Sec. \ref{RelWPS}, it follows from Eq. (\ref{relpsif}) that even if 
the function $f(\theta)$ describes a broad 
angular distribution, the star will be visible only in the angular range of the order of
$R/L$ where $R$ is the radius of the star and $L$ is the distance to the star.
The experimental verification of this prediction is problematic since the quantities $R/L$ 
are very small and
at present star radii cannot be measured directly. Conclusions about them are made from the data on luminosity and temperature assuming that the major part of the radiation from stars comes not from transitions between atomic levels but 
from processes which can be approximately described as a blackbody radiation.  
 
A theoretical model describing blackbody radiation (see e.g. Ref. \cite{LLV}) is such that photons are treated
as an ideal Bose gas weakly interacting with matter and such that typical photon energies are not close to  
energies of absorption lines for that matter (hence the energy spectrum of photons is almost continuous). 
It is also assumed that the photons are distributed over states with definite values of momenta. With these 
assumptions one can derive the famous
Planck formula for the spectral distribution of the blackbody radiation (this formula is treated as
marking the beginning of quantum theory). As explained in Ref. \cite{LLV}, when the photons leave the black body,
their distribution in the phase space can be described by the Liouville theorem; in particular it implies
that the photons leaving stars are moving along classical trajectories. 

If we accept those arguments then the main part of photons emitted by stars can be described in the formalism 
considered in Sec. \ref{RelWPS}. In that case we cannot estimate the quantity $b$ as above and it is 
not clear what criteria can be used for estimating the quantity $a$. 
The estimation $a\approx b\approx 0.48m$ seems to be
extremely favorable since one might expect that the value of $a$ is of atomic size, i.e. much less than $0.48m$. With this
estimation for yellow light (with $\lambda=580nm$) $N_{\bot}=a/\lambda\approx 8\cdot 10^5$. So the value of $N_{\bot}$ is rather large 
and in view of Eq. (\ref{tv}) one might think that the effect of spreading is not important. 

However, this
is not the case because, as follows from Eq. (\ref{tv}), $t_*\approx 0.008s$. Even in the case of
the Sun the distance to the Earth is approximately $t=8$ light minutes,
 and this time is much greater than $t_*$. Then the value of $a(t)$
(which can be called the half-width of the wave packet) when the packet arrives to the Earth is $v_*t\approx 28km$. 
In this case standard geometrical interpretation does not apply. In addition, if we assume that the initial value of $a$ is of the order of several wave lengths then the value of $N_{\bot}$ is much less and the width of the wave packet 
coming to the Earth even from the Sun is much greater. An analogous estimation shows that even in the favorable scenario the half-width 
of the wave packet coming to the Earth 
from Sirius will be approximately equal to $15\cdot 10^6km$ but in less favorable situations the half-width will
be much greater. Hence we come to the conclusion that even in favorable scenarios the assumption that 
photons are moving along classical trajectories does not
apply and a problem arises whether or not this situation is in agreement with experiment. 

As already noted, even if the function $f(\theta)$ describes a broad 
angular distribution, a star will be visible only in the angular range of the order of
$R/L$. Hence one might think
that the absence of classical trajectories does not contradict observations. We now consider this
problem in greater details. For simplicity we first assume that the photon wave function is spherically symmetric, i.e. $f({\bf r}/r)=const$.  

As follows from Eqs. (\ref{FG}) and (\ref{relpsif}), the wave function of the photon coming to Earth
from a distant star is not negligible only within a narrow sphere with the radius $ct$ and the width
of the order of $b$. On its way to Earth the sphere passes {\it all} stars, planets and other objects
the distance from which to the star is less than $L$ (in particular, even those objects which are from
the star in directions opposite to the direction to Earth). 
A problem arises how to explain the fact that the photon was detected on Earth and
escaped detection by those stars, planets etc. 

One might think that the event when the photon was detected on Earth is
purely probabilistic. The fact that the photon was not detected by the objects on its way
to Earth can be explained such that since the photon wave function has a huge size (of the
order of light years or more) the probability of detection even by stars is extremely small
and so it was only a favorable accident that the photon was detected on Earth.

If we accept this explanation then a new problem arises. If the photon passed stars, planets
and other objects on its way to Earth then with approximately the same probability it can
pass Earth and can be detected on the opposite side of the Earth. In that case we could 
see stars even through the Earth.

Moreover, consider the following experiment. Suppose that we first look at a star and then place 
a small screen between the eye and the star. Then the experiment shows that the star will not
be visible. However, since the photon wave function passed many big objects without
interacting with them then with approximately the same probability it can pass the screen.
In that case we could see the star through the screen. 

Another possibility is to try to avoid the above paradoxes by using an 
analogy with classical diffraction
theory.  Here the general problem statement requires solutions of Maxwell's equations with boundary conditions
depending on the shape of the body and its material. In practice this problem is tackled 
assuming that deviation from geometrical optics is small (see e.g. Ref. \cite{LLII}). 
When a classical wave encounters a macroscopic object it is also assumed that in optical phenomena
the wave cannot penetrate inside the object. Then we get a picture that 
 the wave far from the object does not change, right after the object the wave has
a hole but when the length is much greater than the Rayleigh one the hole disappears and
the wave function is practically the same as without diffration. Those results 
are natural from the point of view
that classical waves consist of many almost pointlike photons. 

Let us now consider a single-photon experiment on the Earth such the photon 
encounters a classical object and the transversal width of the photon coordinate wave
function is much greater than the size of the object. One might think that the 
classical diffraction theory can be used even in this case. The justification
involves arguments similar to those in Dirac's textbook \cite{Dirac} and in Sec. \ref{geom} 
that in some cases the classical and quantum
theories involve the same formulas but they have different interpretations.
Then the behavior of the photon wave function after passing the object will be
similar to the behavior of the wave in classical diffraction theory. 

However, any change of the photon transverse wave function implies that the photon 
somehow interacted with the object. For example, when the photon is absorbed by an atom
and then reemitted, the size of its wave function is defined by the atom; so the photon
will not have a broad wave function anymore or in other words the photon
wave function will collapse. Another example is that in the Compton scattering the
photon is first absorbed by a charge particle, in the virtual intermediate state there
is no photon and it is reemitted. So again the wave function will not have a large 
transverse size and the collapse will occur. In general, any Feynman diagram containing photons
consists only of vertices with one photon. So in any interaction the photon will be
first absorbed and hence the reemitted photon will not have the wave function with
a large transverse size. In summary, since the phenomenon of wave function collapse exists only in
quantum theory, in the single-photon experiment discussed above
the behavior of the photon wave after passing the object cannot be similar to the
behavior of the wave in classical diffraction theory.

Let us now return to the case when a photon with a wave function having a cosmic
size encounters an object. In addition to the above arguments one can notice the following.
The assumption that the photon wave function cannot penetrate inside
the macroscopic object is reasonable in experiments on the Earth but in the given
case it is highly problematic.

For example, our understanding of neutrino physics implies that neutrinos not only
can pass the Earth practically without problems but even neutrinos created in the
center of the Sun can easily reach the Earth. The major neutrino detectors are
under the Earth surface and, for example, in the OPERA and ICARUS experiments
neutrinos created at CERN reached Gran Sasso (Italy) after traveling 730km under
the Earth surface. The explanation is that the probability of the neutrino interaction 
with the particle comprising the Sun and the Earth is very small.

At small energies the electromagnetic interaction is much stronger than the weak one but,
as follows from the discussion in Secs. \ref{Mott} and \ref{RelWPS}, the probability of interaction 
for photons having cosmic sizes contains the factor $|{\tilde f}/f|^2=(d/D)^2$. 
Therefore it is reasonable to expect that for such photons the 
probability of interaction with particles comprising an object is even much less
than in the above experiments with neutrinos. Hence the requirement
that the photon wave function cannot penetrate a classical object is not justified.
In addition, by analogy with the above consideration, after every interaction of
the photon with particles comprising the object the photon wave function will collapse
and will not have a cosmic size anymore.  Such a photon can reach Earth only if its momentum 
considerably differs from the original one but this contradicts Theory A.
So the assumption that the above paradoxes can be explained by analogy with classical 
diffraction theory is not justified. 

If $f({\bf r}/r)\neq const$ then, as follows from Eqs. (\ref{FG}) and (\ref{relpsif}),
the radial part of the wave function is the same as in the spherically symmetric case and, 
as follows from the above discussion, the photon coordinate wave function still has a cosmic size.
Therefore on its way to Earth the photon wave function will also pass stars, planets and
other objects (even if they are far from the line connecting the star and Earth) 
and the same inconsistencies arise.

{\it In summary, since according to standard theory photons emitted by stars have coordinate wave functions with cosmic sizes,
we treat the above arguments as a strong indication that the theory contradicts observational data.}

In the infrared and radio astronomy wave lengths are much greater than in the optical region but typical values
of $a_{ph}$ are expected to be much greater. As a consequence, here standard quantum theory
encounters the same problems that in the optical region.

In the case of gamma-ray bursts (GRBs) wave lengths are much less than in the optical region but this is 
outweighed by the facts that, according
to the present understanding of the GRB phenomenon (see e.g. Ref. \cite{bursts}), gamma quanta created in GRBs 
typically travel to Earth for billions of years and typical values
of $a_{ph}$ are expected to be much less than in the optical region. The location of sources of GBRs 
are determined with a good
accuracy and the data can be explained only assuming that the gamma quanta are focused into  
narrow jets 
which are observable when Earth lies along the path of those jets. However, in view of the above
discussion, the results on WPS predicted by standard quantum theory are incompatible with the
data on GRBs because, as a consequence of WPS, the probability to detect photons from
GRBs would be negligible. 

Consider now WPS effects for radio wave photons. In radiolocation it is important
that a beam from a directional antenna has a narrow angular distribution and a narrow distribution of wave lengths.
 This makes it possible to communicate even with very distant 
space probes. For this purpose a set of radio telescopes can be used but for simplicity we consider
a model where signals from a space probe are received by one radio telescope having the diameter $D$ of the dish.

The Cassini spacecraft can transmit to Earth at three radio wavelengths: 14cm, 4cm and 1cm \cite{Cassini}.
A radio telescope on Earth can determine the position of Cassini with a good accuracy if it detects photons having
momenta in the angular range of the order of $D/L$ where $L$ is the
distance to Cassini. The main idea of using a system of radio telescopes is to increase the effective value of $D$. 
As a consequence of the fact that the radio signal sent from Cassini has an angular divergence
which is much greater than $D/L$, only a small part of photons in the signal can be detected. 
We consider a case when Cassini was 7AU away from the Earth.

Consider first the problem on classical level. For the quantity $a=a_{cl}$ we take the value of $1m$
which is of the order of the radius of the Cassini antenna. If $\alpha=\lambda/(2\pi a)$ and $L(t)$
is the length of the classical path then, as follows from Eq. (\ref{em}), $a_{cl}(t)\approx L(t)\alpha$. As a 
result, even for $\lambda=1cm$ we have $a_{cl}(t)\approx 1.6\cdot 10^6km$. Hence one 
might expect that only
a $[D/a_{cl}(t)]^2$ part of the photons can be detected. 

Consider now the problem on quantum level. The condition
$t\gg t_*$ is satisfied for both, the classical and quantum problems. Then, as follows from Eq. (\ref{em}), $a_{ph}(t)=a_{cl}(t)a_{cl}/a_{ph}$,
i.e. the quantity $a_{ph}(t)$ is typically greater than $a_{cl}(t)$ and in Sec. \ref{WPW} this effect is called
the WPW paradox. The fact that only photons in the angular range $D/L$ can be detected can be described by
projecting the states $\chi=\chi({\bf p},t)$ (see Eqs. (\ref{chiprel}), and (\ref{chiptphoton})) onto the states
$\chi_1={\cal P}\chi$ where $\chi_1({\bf p},t)=\rho({\bf p})\chi({\bf p},t)$ and the form factor $\rho({\bf p})$ is
significant only if ${\bf p}$ is in the needed angular range. We choose 
$\rho({\bf p})=exp(-{\bf p}_{\bot}^2a_1^2/2\hbar^2)$ where $a_1$ is of the order of $\hbar L/(p_0 D)$. 
Since $a_1\gg a_{ph}$,
it follows from Eqs. (\ref{chiprel}), and (\ref{chiptphoton}) that $||{\cal P}\chi||^2=(a_{ph}/a_1)^2$.  Then, as follows from Eq. (\ref{em}),
$(a_{ph}/a_1)^2$ is of the order of $[D/a_{ph}(t)]^2$ as expected and this quantity is typically much less than 
$[D/a_{cl}(t)]^2$. Hence the WPW paradox would make communications with space probes 
much more difficult. 

We now consider the following problem. The parameter $\gamma$ in General Relativity (GR) 
is extracted from experiments on deflection of light from distant stars by the Sun and from the effect
called Shapiro time delay. The meaning of the effect follows. An antenna
on Earth sends a signal to Mercury, Venus or an interplanetary space probe and receives the reflected signal.
If the path of the signal nearly grazes the Sun then the gravitational influence of the Sun deflects the path
from a straight line. As a result, the path becomes longer by 
$S\approx 75km$ and the signals arrive with a
delay $S/c\approx 250\mu s$. This effect is treated as the fourth test of GR.

The consideration of the both effects in GR is based on the assumption that the photon 
is a pointlike classical particle moving along classical trajectory. In the first case the
photon wave function has a cosmic size. In the second case the available experimental data are treated
such that the best test of $\gamma$ has been performed in measuring the Shapiro delay when signals from the DSS-25 antenna \cite{DSS25} were sent to the Cassini spacecraft when 
it was 7AU away from the Earth. As noted above, in that case case, even in the most favorable 
scenario $a_{cl}(t)\approx 1.6\cdot 10^6km$ and the quantity $a_{ph}(t)$ is expected to be
much greater. Therefore a problem arises whether the classical consideration in GR is
compatible with the fact that the photon coordinate wave functions have very large sizes.

One might think that the compatibility is not a problem because when we detect a photon with the momentum 
pointing to the area near the Sun
we know that this photon moved to us on the trajectory bending near the Sun.
The results of Sec. \ref{RelWPS} indeed show that even if the photon momentum wave function
has a broad distribution, the photon detected by a measuring device can be detected only
at the moment of time close to $L/c$ and momentum of the detected photon will point to the star which
emitted this photon. However, quantum formalism does not contain any information about the
photon trajectory from the moment of emission to the moment of detection. One might 
guess that the required trajectory will give the main contribution in the Feynman path integral 
formulation but the proof of this guess is rather complicated.

In summary, by analogy with the consideration in Subsec. \ref{when}, one can conclude 
that quantum theory does not contain any information about trajectories. The notion of
trajectories in quantum theory is a reasonable approximation only in semiclassical 
approximation when a choice of the position operator has been made. However, in
the case of packets with broad coordinate distributions the notion of trajectories
does not have a physical meaning and one cannot avoid quantum consideration of
the problem. In particular, the results of GR on the deflection of light and on the Shapiro
delay are meaningful only
if there is no considerable WPS in quantum theory. In addition, in view of the WPW
paradox, the probability to detect reflected photons in the Shapiro delay experiments can be very small.

One might think that the  WPS effect is
important only if a particle travels a rather long distance. Hence one might expect that in experiments 
on the Earth this effect is negligible.
Indeed, one might expect that in typical experiments on the Earth the time $t$ is so small that $a(t)$ 
is much less than the size of any macroscopic source of light. However, a conclusion that 
the effect of WPS is
negligible for any experiment on the Earth might be premature.

As an example, consider the case of protons in the LHC accelerator. According to Ref. \cite{protons},
protons in the LHC ring injected at the energy $E=450 GeV$ should be accelerated to the energy $E=7 TeV$ within
one minute during which the protons will turn around the $27km$ ring approximately 674729 times. Hence the length
of the proton path is of the order of $18\cdot 10^6km$. The protons cannot be treated as free particles since
they are accelerated by strong magnets. A problem of how the width of the proton wave function behaves in the
presence of strong electromagnetic field is very complicated and the solution of the problem is not known yet.
It is always assumed that the WPS effect for the protons can be neglected. 

We first consider a model problem of
the WPS for a free proton which moves for $t_1=1min$ with the energy in the range $[0.45,7]\, TeV$.
In nuclear physics the size of the proton is usually assumed to be a quantity of the order of $10^{-13}cm$. 
Therefore for estimations we take $a=10^{-13}cm$. Then the quantity $t_*$ defined after Eq. (\ref{at}) is not
greater than $10^{-19}s$, i.e. $t_*\ll t_1$. Hence, as follows from Eq. (\ref{at}), the quantity 
$a(t_1)$ is of the order of $500km$ if $E=7\, TeV$ and  by a factor of
$7/0.45\approx 15.6$ greater if $E=450\, GeV$. 

This fully unrealistic result cannot be treated as a paradox since, as noted
above, the protons in the LHC ring are not free. In the real situation the protons interact
with many real and virtual photons emitted by magnets. For example, this might lead to
the collaps of the proton wave function each time when the proton interacts with the
real or virtual photon. This phenomenon is not well studied yet and so  
a problem of what standard theory 
predicts on the width of proton wave functions in the LHC ring is far from being obvious.

The last example follows. The astronomical objects called pulsars are treated such that they are neutron stars
with radii much less than radii of ordinary stars. Therefore if mechanisms of pulsar electromagnetic radiation
were the same as for ordinary stars then the pulsars would not be visible. The fact that pulsars are visible is
explained as a consequence of the fact that they emit beams of
light which can only be seen when the light is pointed in the direction of the observer with some periods which 
are treated
as periods of rotation of the neutron stars. In popular literature this is compared with the light of a lighthouse.
However, by analogy with the case of a signal sent from Cassini, only a small part of photons in the beam can reach the Earth.
At present the pulsars have been observed in different regions of the electromagnetic spectrum but the first
pulsar called PSR B1919+21 was discovered in 1967 as a radio wave radiation with $\lambda\approx 3.7m$ \cite{pulsar}.
This pulsar is treated as the neutron star with the radius $R=0.97km$ and the distance from the pulsar to
the Earth is 2283 light years. If for estimating $a_{cl}(t)$ we assume that   
$a_{cl}=R$ then we get $\alpha\approx 6\cdot 10^{-4}$ and $a_{cl}(t)\approx 1.3 ly\approx 12\cdot 10^{12}km$.
Such an extremely large value of spreading poses a problem whether even predictions of classical electrodynamics are
compatible with the fact that pulsars are observable. However, in view of the WPW paradox, 
the value of $a_{ph}(t)$ will be even much greater and no observation of pulsars would be possible.

Our conclusion is that we have several fundamental paradoxes indicating that predictions of standard 
quantum theory for the WPS effect contradict experimental data. 

\section{Discussion: is it possible to avoid the WPS paradoxes in standard theory?}
\label{Discussion}

As shown in the preceding section, if one assumes that photons coming to Earth do not interact 
with the interstellar or interplanetary medium and with each other then a standard treatment of the WPS effect leads to several paradoxes. Hence a question arises whether this assumption is legitimate. 

As shown in textbooks on quantum optics (see e.g. Refs. \cite{Mandel,Scully}), quantum states
describing the laser emission are strongly coherent and the approximation
of independent photons is not legitimate. As shown in Sec. \ref{coherent}, the WPS effect in coherent states
is pronounced in a much less extent than for individual photons. 
However, laser emission can be created only at very special conditions
when energy levels are inverted, the emission is amplified in the laser cavity etc. At the same time, the
main part of the radiation emitted by stars is understood such that it can be approximately described as 
the blackbody radiation and in addition a part of the radiation consists of photons emitted from different atomic
energy levels. In that case the emission of photons is spontaneous rather than induced and one might think that 
the photons can be treated independently. Several authors (see e.g. Ref. \cite{Letokhov} and references therein)
discussed a possibility that at some conditions the inverted population and amplification of radiation in stellar
atmospheres might occur and so a part of the radiation can be induced. This problem is now under investigation. 
Hence we adopt a standard assumption that a main part of the radiation from stars is spontaneous. In addition, 
there is no reason to think that radiation of GRBs, radio antennas, space probes or pulsars is laser like. 

The next question is whether the interaction of photons in the above phenomena is important or not.
As explained in standard textbooks on QED (see e.g. Ref. \cite{AB}), the photon-photon
interaction can go only via intermediate creation of virtual electron-positron or quark-antiquark pairs. If $\omega$ is
the photon frequency, $m$ is the mass of the charged particle in the intermediate state and $e$ is the electric charge
of this particle then in the case
when $\hbar\omega\ll mc^2$ the total cross section of the photon-photon interaction is \cite{AB}
\begin{equation}
\sigma=\frac{56}{5\pi m^2}\frac{139}{90^2}(\frac{e^2}{\hbar c})^4 (\frac{\hbar\omega}{mc^2})^6
\label{photonphoton}
\end{equation} 
For photons of visible light the quantities $\hbar\omega/(mc^2)$ and $\sigma$ are very small and for radio waves
they are even smaller by several orders of magnitude. At present the effect of the direct
photon-photon interaction has not been detected, and experiments with strong laser fields were only able to 
determine the upper limit of the cross section \cite{gammagamma}.

The problem of WPS in the ultrarelativistic case has been discussed
in a wide literature. As already noted, in Ref. \cite{Dillon} the effect 
of WPS has been discussed in the Fresnel approximation for a two-dimensional model and the author shows that
in the direction perpendicular to the group velocity of the wave spreading is important. He considers WPS in
the framework of classical electrodynamics. We believe that considering this effect from quantum point of view
is even simpler since the photon wave function satisfies the relativistic Schr\"{o}dinger equation which is linear
in $\partial /\partial t$. As noted in Sec. \ref{geom}, this function also satisfies the wave equation but
it is simpler to consider an equation linear in $\partial /\partial t$ than that quadratic in
$\partial /\partial t$. However, in classical theory there is no such an object as the photon wave function and
hence one has to solve either a system of Maxwell equations or the wave equation. 
There is also a number of works where the authors consider WPS in view of propagation of classical waves in 
a medium such that dissipation is important (see e.g. Ref. \cite{Christov}). In Ref. \cite{Wang}
the effect of WPS has been discussed in view of a possible existence of superluminal neutrinos. The authors
consider only the dynamics of the wave packet in the longitudinal direction in the framework of the
Dirac equation. They conclude that wave packets describing ultrarelativistic fermions do not experience WPS in this direction. However, the authors do not consider WPS in perpendicular directions.

In view of the above discussion, standard treatment of WPS leads to several fundamental paradoxes.
To the best of our knowledge, those paradoxes have never been discussed in the literature. For resolving the
paradoxes one could discuss several possibilities. One of them might be 
such that the interaction of light with the interstellar or interplanetary medium cannot be neglected.
On quantum level a process of propagation of photons in the medium is rather complicated because several
mechanisms of propagation should be taken into account. For example, a possible process is such that a
photon can be absorbed by an atom and reemitted. This process makes it
clear why the speed of light in the medium is less than $c$: because the atom which absorbed the photon is
in an excited state for some time before reemitting the photon. However, this process is also important from the
following point of view: even if the coordinate photon wave function had a large width before absorption,
as a consequence of the collapse of the wave function, the wave function of the emitted
photon will have in general much smaller dimensions since after detection the width is
defined only by parameters of the corresponding detector. If the photon encounters many atoms on its way,
this process does not allow the photon wave function to spread out significantly. Analogous remarks can
be made about other processes, for example about rescattering of photons on large groups of atoms, rescattering
on elementary particles if they are present in the medium etc. However, such processes have been discussed in
Theory B and, as noted in Sec. \ref{experiment}, they probably result in more blurring than is seen. 

The interaction of photons with the interstellar or interplanetary medium might also be important in view of
hypotheses that the density of the medium is much greater than usually believed. Among the most popular
scenarios are dark energy, dark matter etc. As shown in our papers (see e.g. Refs. \cite{symm1401,DS,dark} and references
therein), the phenomenon of the cosmological acceleration can be easily and naturally explained from first
principles of quantum theory without involving dark energy, empty space-background and other artificial notions. 
However, the other scenarios seem to be more realistic and one might expect that they will be intensively investigated. A rather hypothetical
possibility is that the propagation of photons in the medium has something in common with
the induced emission when a photon induces emission of other photons in practically the same direction.
In other words, the interstellar medium amplifies the emission as a laser. This possibility seems to be not 
realistic since it is not clear why the energy levels in the medium might be inverted.  

We conclude that at present in standard theory there are no realistic scenarios which can explain
the WPS paradoxes. In the remaining part of the paper we propose a solution of the problem proceeding from a consistent definition of the position operator. 

\section{Consistent construction of position operator}
\label{consistent}

The above results give grounds to think that the reason of the paradoxes which follow from the behavior of the 
coordinate photon wave function in perpendicular directions is that standard definition of the position operator in 
those directions does not correspond to realistic measurements of coordinates. 
Before discussing a consistent construction, let us make the following
remark. On elementary level students treat the mass $m$ and the velocity ${\bf v}$ as primary quantities such
that the momentum is $m{\bf v}$ and the kinetic energy is $m{\bf v}^2/2$. However, from the point of view of
Special Relativity, the primary quantities are the momentum ${\bf p}$ and the total energy $E$ and then the mass
and velocity are defined as $m^2c^4=E^2-{\bf p}^2c^2$ and ${\bf v}={\bf p}c^2/E$, respectively. This example has
the following analogy. In standard quantum theory the primary operators are the position and momentum operators
and the orbital angular momentum operator is defined as their cross product. However,  
the operators ${\bf P}$ and ${\bf L}$ are consistently defined as representation operators of the Poincare algebra
while the definition of the position operator is a problem. Hence a question arises whether the position
operator can be defined in terms of ${\bf P}$ and ${\bf L}$.

One might seek the position operator such that on classical level
the relation ${\bf r}\times{\bf p}={\bf L}$ will take place. Note that on quantum level this relation is not
necessary. Indeed, the very fact that some elementary particles have a half-integer spin
shows that the total angular momentum for those particles does not have the orbital nature but on classical 
level the angular momentum can be always represented as a cross
product of the radius-vector and standard momentum. However, if the values 
of ${\bf p}$ and ${\bf L}$ are
known and ${\bf p}\neq 0$ then the requirement that ${\bf r}\times{\bf p}={\bf L}$ does not define ${\bf r}$ 
uniquely. One can define parallel and perpendicular components of ${\bf r}$ as 
${\bf r}=r_{||}{\bf p}/p+{\bf r}_{\bot}$ where $p=|{\bf p}|$. Then the relation ${\bf r}\times{\bf p}={\bf L}$ 
defines uniquely only ${\bf r}_{\bot}$. Namely, as follows from this relation, 
${\bf r}_{\bot}=({\bf p}\times{\bf L})/p^2$. In view of the fact that on quantum level the operators
${\bf p}$ and ${\bf L}$ do not commute, on this level  
${\bf r}_{\bot}$ should be replaced by a selfadjoint operator 
${\bf {\cal R}}_{\bot}=({\bf p}\times{\bf L}-{\bf L}\times{\bf p})/(2p^2)$. Therefore 
\begin{eqnarray}
&&{\cal R}_{\bot j}=\frac{\hbar}{2p^2}e_{jkl}(p_kL_l+L_lp_k)=\frac{\hbar}{p^2}e_{jkl}p_kL_l-
\frac{i\hbar}{p^2}p_j\nonumber\\
&&=i\hbar\frac{\partial}{\partial p_j}-i\frac{\hbar}{p^2}p_jp_k\frac{\partial}{\partial p_k}-\frac{i\hbar}{p^2}p_j
\label{rbot}
\end{eqnarray}
where $e_{jkl}$ is the absolutely antisymmetric tensor, $e_{123}=1$, a sum over repeated indices is assumed and 
we assume that if ${\bf L}$ is given by Eq. (\ref{IRoperators}) then the orbital momentum is $\hbar{\bf L}$.

We define the operators ${\bf F}$ and ${\bf G}$ such that ${\bf {\cal R}}_{\bot}=\hbar{\bf F}/p$ and
${\bf G}$ is the operator of multiplication by the unit vector ${\bf n}={\bf p}/p$.
A direct calculation shows that these operators satisfy the following relations:
\begin{eqnarray}
&&[L_j,F_k]=ie_{jkl}F_l,\quad [L_j,G_k]=ie_{jkl}F_l,\quad {\bf G}^2=1,\quad {\bf F}^2={\bf L}^2+1 \nonumber\\
&&[G_j,G_k]=0,\quad [F_j,F_k]=-ie_{jkl}L_l\quad e_{jkl}\{F_k,G_l\}=2L_j\nonumber\\
&&{\bf L}{\bf G}={\bf G}{\bf L}={\bf L}{\bf F}={\bf F}{\bf L}=0, \quad {\bf F}{\bf G}=-{\bf G}{\bf F}=i
\label{vectorFG}
\end{eqnarray}
The first two relations show that ${\bf F}$ 
and ${\bf G}$ are the vector operators as expected. The result for the anticommutator shows
that on classical level ${\bf F}\times {\bf G}={\bf L}$ and the last two relations show that on classical level
the operators in the triplet $({\bf F},{\bf G},{\bf L})$ are mutually orthogonal. 

Note that if the momentum distribution is narrow and such that the mean
value of the momentum is directed along the $z$ axis then it does not mean that on the operator level the $z$
component of the operator ${\bf {\cal R}}_{\bot}$ should be zero. The matter is that the direction of the momentum
does not have a definite value. One might expect that only the mean value of the operator ${\bf {\cal R}}_{\bot}$
will be zero or very small.

In addition, an immediate consequence of the definition (\ref{rbot}) follows: {\it Since the momentum and
angular momentum operators commute with the Hamiltonian, the distribution of all the components of ${\bf r}_{\bot}$
does not depend on time. In particular, there is no WPS in directions defined by ${\bf {\cal R}}_{\bot}$.}
This is also clear from the fact that ${\bf {\cal R}}_{\bot}=\hbar{\bf F}/p$ where the operator ${\bf F}$ acts only
over angular variables and the Hamiltonian depends only on $p$.
On classical level the conservation of ${\bf {\cal R}}_{\bot}$ is obvious since it is defined by the conserving
quantities ${\bf p}$ and ${\bf L}$. It is also obvious that since a free particle is moving along a 
straight line, a vector from the origin perpendicular to this line does not change with time.

The above definition of the perpendicular component of the position operator is well substantiated
since on classical level the relation ${\bf r}\times{\bf p}={\bf L}$ has been verified in numerous experiments.
However, this relation does not make it possible to define the parallel component of the position operator
and a problem arises what physical arguments should be used for that purpose. 

A direct calculation shows that if $\partial/\partial {\bf p}$ is written in terms of $p$ and angular variables then
\begin{equation}
i\hbar\frac{\partial}{\partial {\bf p}}={\bf G}{\cal R}_{||}+{\bf {\cal R}}_{\bot}
\label{decomp}
\end{equation}
where the operator ${\cal R}_{||}$ acts only over the variable $p$:
\begin{equation}
{\cal R}_{||}=i\hbar (\frac{\partial}{\partial p}+\frac{1}{p})
\label{rparall}
\end{equation}
The correction $1/p$ is related to the fact that the operator ${\cal R}_{||}$ is Hermitian since in variables 
$(p,{\bf n})$ the scalar product is given by
\begin{equation}
(\chi_2,\chi_1)=\int \chi_2(p,{\bf n})^*\chi_1(p,{\bf n})p^2dp do
\label{scalar}
\end{equation}
where $do$ is the element of the solid angle.

While the components of standard position operator commute
with each other, the operators ${\cal R}_{||}$ and ${\bf {\cal R}}_{\bot}$ satisfy the following commutation relations:
\begin{equation}
[{\cal R}_{||},{\bf {\cal R}}_{\bot}]=-\frac{i\hbar}{p}{\bf {\cal R}}_{\bot},\quad 
[{\cal R}_{\bot j},{\cal R}_{\bot k}]
=-\frac{i\hbar^2}{p^2}e_{jkl}L_l
\label{rparallrbot}
\end{equation}
An immediate consequence of these relations follows: {\it Since the operator ${\cal R}_{||}$ and different
components of ${\bf {\cal R}}_{\bot}$ do not commute with each other, the corresponding quantities cannot be
simultaneously measured and hence there is no wave function $\psi(r_{||},{\bf r}_{\bot})$ in 
coordinate representation.}

In standard theory $-\hbar^2 (\partial/\partial{\bf p})^2$ is the operator of the quantity ${\bf r}^2$. As follows
from Eq. (\ref{vectorFG}), the two terms in Eq. (\ref{decomp}) are not strictly orthogonal and on the operator level
$-\hbar^2 (\partial/\partial{\bf p})^2\neq {\cal R}_{||}^2+{\bf {\cal R}}_{\bot}^2$.
A direct calculation using Eqs. (\ref{vectorFG}) and (\ref{decomp}) gives
\begin{equation}
\frac{\partial^2}{\partial{\bf p}^2}=\frac{\partial^2}{\partial p^2}+\frac{2}{p}\frac{\partial}{\partial p}-
\frac{{\bf L}^2}{p^2},\quad -\hbar^2\frac{\partial^2}{\partial{\bf p}^2}={\cal R}_{||}^2+{\bf {\cal R}}_{\bot}^2-
\frac{\hbar^2}{p^2}
\label{r2}
\end{equation}
in agreement with the expression for the Laplacian in spherical coordinates. In semiclassical approximation, 
$(\hbar^2/p^2)\ll {\bf {\cal R}}_{\bot}^2$ since the eigenvalues of ${\bf L}^2$ are $l(l+1)$, 
in semiclassical states $l\gg 1$ and, as follows from Eq. (\ref{vectorFG}),
${\bf {\cal R}}_{\bot}^2=[\hbar^2(l^2+l+1)/p^2]$.

As follows from Eq. (\ref{rparallrbot}), $[{\cal R}_{||},p]=-i\hbar$, i.e. in the longitudinal direction the
commutation relation between the coordinate and momentum is the same as in standard theory. One can also calculate
the commutators between the different components of ${\bf {\cal R}}_{\bot}$ and ${\bf p}$. Those commutators are
not given by such simple expressions as in standard theory but it is easy to see that all of them are of the
order of $\hbar$ as it should be.

Equation (\ref{decomp}) can be treated as an implementation of the relation 
${\bf r}=r_{||}{\bf p}/|{\bf p}|+{\bf r}_{\bot}$ on quantum level. As argued in Secs. \ref{intropos}
and \ref{classical}, standard position operator $i\hbar\partial/\partial p_j$ in the direction $j$ is not consistently
defined if $p_j$ is not sufficiently large. One might think however that since the operator ${\cal R}_{||}$ contains
$i\hbar\partial/\partial p$, it is defined consistently if the magnitude of the momentum is sufficiently large. 

In summary, we propose to define the position operator not by the set $(i\hbar\partial/\partial p_x,
i\hbar\partial/\partial p_y,i\hbar\partial/\partial p_z)$ but by the operators ${\cal R}_{||}$ and ${\bf {\cal R}}_{\bot}$.
Those operators are defined from different considerations. As noted above, the definition of ${\bf {\cal R}}_{\bot}$ is 
based on solid physical facts while the definition of ${\cal R}_{||}$ is expected to be more consistent than the
definition of standard position operator. However, this does not guarantee that the operator ${\cal R}_{||}$
is consistently defined in all situations. As argued in Ref. \cite{gravity}, in a quantum theory over a Galois field
an analogous definition is not consistent {\it for macroscopic bodies} (even if $p$ is large) since in that case 
semiclassical approximation is not valid. In the remaining part of the paper we assume that for elementary particles
the above definition of ${\cal R}_{||}$ is consistent in situations when semiclassical approximation applies.

One might pose the following question. What is the reason to work with the parallel and perpendicular components
of the position operator separately if, according to Eq. (\ref{decomp}), their sum is the standard position operator?
The explanation follows.

In quantum theory every physical quantity corresponds to a selfadjoint operator but the theory does not define
explicitly how a quantity corresponding to a specific operator should be measured. There is no guaranty that for
each selfadjoint operator there exists a physical quantity which can be measured in real experiments.

Suppose that there are three physical quantities corresponding to the selfadjoint operators $A$, $B$ and $C$ such that
$A+B=C$. Then in each state the mean values of the operators are related as ${\bar A}+{\bar B}={\bar C}$ but in situations 
when the operators $A$ and $B$ do not commute with each other there is no direct relation between the distributions
of the  physical quantities corresponding to the operators $A$, $B$ and $C$. For example, in situations when the physical 
quantities corresponding to the
operators $A$ and $B$ are semiclassical and can be measured with a good accuracy, there is no guaranty that the physical 
quantity corresponding to the operator $C$ can be measured in real measurements. As an example, the physical meaning
of the quantity corresponding to the operator $L_x+L_y$ is problematic. Another example is the situation with WPS in 
directions perpendicular to the particle momentum. Indeed, as noted above, the physical
quantity corresponding to the operator ${\bf {\cal R}}_{\bot}$ does not experience WPS and, as shown in Sec. \ref{newWPS},
in the case of ultrarelativistic particles there is no WPS in the parallel direction as well. However,  
standard position operator is a sum of noncommuting operators corresponding to well defined physical
quantities and, as a consequence, there are situations when standard position operator defines a quantity which
cannot be measured in real experiments.

\section{New position operator and semiclassical states}
\label{newsemicl}

As noted in Sec. \ref{classical}, in standard theory states are treated as semiclassical in greatest possible extent
if $\Delta r_j \Delta p_j =\hbar/2$ for each $j$ and such states are called coherent. The existence of coherent
states in standard theory is a consequence of commutation relations $[p_j,r_k]=-i\hbar \delta_{jk}$. Since in our
approach there are no such relations, a problem arises how to construct states in which all physical quantities
$p$, $r_{||}$, ${\bf n}$ and ${\bf r}_{\bot}$ are semiclassical.

One can calculate the mean values and uncertainties of the operator ${\cal R}_{||}$ and
all the components of the operator 
${\bf {\cal R}}_{\bot}$ in the state defined by Eq. (\ref{chiprel}). The calculation is not simple since it involves
three-dimensional integrals with Gaussian functions divided by $p^2$. The result is that these operators
are semiclassical in the state (\ref{chiprel}) if $p_0\gg \hbar/b$, $p_0\gg \hbar/a$ and 
$r_{0z}$ has the same order of magnitude as $r_{0x}$ and $r_{0y}$. 

However, a more natural approach follows. Since ${\bf {\cal R}}_{\bot}=\hbar{\bf F}/p$, the operator
${\bf F}$ acts only over the angular
variable ${\bf n}$ and ${\cal R}_{||}$ acts only over the variable $p$, it is convenient to work
in the representation where the Hilbert space is the space of functions $\chi(p,l,\mu)$ such that 
the scalar product is
\begin{equation}
(\chi_2,\chi_1)=\sum_{l\mu}\int_0^{\infty} \chi_2(p,l,\mu)^*\chi_1(p,l,\mu)dp
\label{newscalar}
\end{equation} 
and $l$ and $\mu$ are the orbital and magnetic quantum numbers, respectively, i.e.
\begin{equation}
{\bf L}^2\chi(p,l,\mu)=l(l+1)\chi(p,l,\mu),\quad L_z\chi(p,l,\mu)=\mu\chi(p,l,\mu)
\label{orbmagn}
\end{equation}

The operator ${\bf L}$ in this space does not act over the variable $p$ and the action of the remaining
components is
given by
\begin{equation}
L_+\chi(l,\mu)=[(l+\mu)(l+1-\mu)]^{1/2}\chi(l,\mu-1),\quad
L_-\chi(l,\mu)=[(l-\mu)(l+1+\mu)]^{1/2}\chi(l,\mu+1)
\label{L}
\end{equation}
where the $\pm$ components of vectors are defined such that $L_x=L_++L_-$, $L_y=-i(L_+-L_-)$.

A direct calculation shows that, as a consequence of Eq. (\ref{rbot})
\begin{eqnarray}
&&F_+\chi(l,\mu)=-\frac{i}{2}[\frac{(l+\mu)(l+\mu-1)}{(2l-1)(2l+1)}]^{1/2}l\chi(l-1,\mu-1)\nonumber\\
&&-\frac{i}{2}[\frac{(l+2-\mu)(l+1-\mu)}{(2l+1)(2l+3)}]^{1/2}(l+1)\chi(l+1,\mu-1)\nonumber\\
&&F_-\chi(l,\mu)=\frac{i}{2}[\frac{(l-\mu)(l-\mu-1)}{(2l-1)(2l+1)}]^{1/2}l\chi(l-1,\mu+1)\nonumber\\
&&+\frac{i}{2}[\frac{(l+2+\mu)(l+1+\mu)}{(2l+1)(2l+3)}]^{1/2}(l+1)\chi(l+1,\mu+1)\nonumber\\
&&F_z\chi(l,\mu)=i[\frac{(l-\mu)(l+\mu)}{(2l-1)(2l+1)}]^{1/2}l\chi(l-1,\mu)\nonumber\\
&&-i[\frac{(l+1-\mu)(l+1+\mu)}{(2l+1)(2l+3)}]^{1/2}(l+1)\chi(l+1,\mu)
\label{F}
\end{eqnarray}
The operator ${\bf G}$ acts on such states as follows
\begin{eqnarray}
&&G_+\chi(l,\mu)=\frac{1}{2}[\frac{(l+\mu)(l+\mu-1)}{(2l-1)(2l+1)}]^{1/2}\chi(l-1,\mu-1)\nonumber\\
&&-\frac{1}{2}[\frac{(l+2-\mu)(l+1-\mu)}{(2l+1)(2l+3)}]^{1/2}\chi(l+1,\mu-1)\nonumber\\
&&G_-\chi(l,\mu)=-\frac{1}{2}[\frac{(l-\mu)(l-\mu-1)}{(2l-1)(2l+1)}]^{1/2}\chi(l-1,\mu+1)\nonumber\\
&&+\frac{1}{2}[\frac{(l+2+\mu)(l+1+\mu)}{(2l+1)(2l+3)}]^{1/2}\chi(l+1,\mu+1)\nonumber\\
&&G_z\chi(l,\mu)=-[\frac{(l-\mu)(l+\mu)}{(2l-1)(2l+1)}]^{1/2}\chi(l-1,\mu)\nonumber\\
&&-[\frac{(l+1-\mu)(l+1+\mu)}{(2l+1)(2l+3)}]^{1/2}\chi(l+1,\mu)
\label{G}
\end{eqnarray}
and now the operator ${\cal R}_{||}$ has a familiar form ${\cal R}_{||}=i\hbar \partial/\partial p$. 

Therefore by analogy with Secs. \ref{classical} and \ref{NRWPS} one can construct states which are coherent 
with respect to $(r_{||},p)$, i.e. such that 
$\Delta r_{||}\Delta p=\hbar/2$. Indeed (see Eq. (\ref{chip})), the wave function
\begin{equation}
\chi(p)=\frac{b^{1/2}}{\pi^{1/4}\hbar^{1/2}}exp[-\frac{(p-p_0)^2b^2}{2\hbar^2}-
\frac{i}{\hbar}(p-p_0)r_0]
\label{pr||}
\end{equation}
describes a state where the mean values of $p$ and $r_{||}$ are $p_0$ and $r_0$, respectively and their uncertainties
are $\hbar /(b\sqrt{2})$ and $b/\sqrt{2}$, respectively. Strictly speaking, the analogy between the given case and
that discussed in Secs. \ref{classical} and \ref{NRWPS} is not full since in the given case the quantity $p$ can be
in the range $[0,\infty)$, not in $(-\infty,\infty)$ as momentum variables used in those sections. However, if 
$p_0b/\hbar \gg 1$ then the formal expression for $\chi(p)$ at $p<0$ is extremely small and so the normalization
integral for $\chi(p)$ can be formally taken from $-\infty$ to $\infty$.

In such an approximation one can define wave functions $\psi(r)$ in the $r_{||}$ representation.
By analogy with the consideration in Secs. \ref{classical} and \ref{NRWPS} we define 
\begin{equation}
\psi(r)=\int exp(\frac{i}{\hbar}pr)\chi(p)\frac{dp}{(2\pi\hbar)^{1/2}}
\label{psir||in}
\end{equation}
where the integral is formally taken from $-\infty$ to $\infty$. Then 
\begin{equation}
\psi(r)=\frac{1}{\pi^{1/4}b^{1/2}}exp[-\frac{(r-r_0)^2}{2b^2}+\frac{i}{\hbar}p_0r]
\label{psir||B}
\end{equation}
Note that here the quantities $r$ and $r_0$ have the meaning of coordinates in the direction parallel to the
particle momentum, i.e. they can be positive or negative.

Consider now states where the quantities ${\bf F}$ and ${\bf G}$ are semiclassical.
One might expect that in semiclassical states the quantities $l$ and $\mu$ are very
large. In this approximation, as follows from Eqs. (\ref{F}) and (\ref{G}), the action of 
the operators ${\bf F}$ and ${\bf G}$ can be written as
\begin{eqnarray}
&&F_+\chi(l,\mu)=-\frac{i}{4}(l+\mu)\chi(l-1,\mu-1)-\frac{i}{4}(l-\mu)\chi(l+1,\mu-1)\nonumber\\
&&F_-\chi(l,\mu)=\frac{i}{4}(l-\mu)\chi(l-1,\mu+1)+\frac{i}{4}(l+\mu)\chi(l+1,\mu+1)\nonumber\\
&&F_z\chi(l,\mu)=-\frac{i}{2l}(l^2-\mu^2)^{1/2}[\chi(l+1,\mu)+\chi(l-1,\mu)]\nonumber\\
&&G_+\chi(l,\mu)=\frac{l+\mu}{4l}\chi(l-1,\mu-1)-\frac{l-\mu}{4l}\chi(l+1,\mu-1)\nonumber\\
&&G_-\chi(l,\mu)=-\frac{l-\mu}{4l}\chi(l-1,\mu+1)+\frac{l+\mu}{4l}\chi(l+1,\mu+1)\nonumber\\
&&G_z\chi(l,\mu)=-\frac{1}{2l}(l^2-\mu^2)^{1/2}[\chi(l+1,\mu)+\chi(l-1,\mu)]
\label{FGsemicl}
\end{eqnarray}

In view of the remark in Sec. \ref{classical} about semiclassical vector quantities, 
consider a state $\chi(l,\mu)$ such that $\chi(l,\mu)\neq 0$ only if $l\in [l_1,l_2]$, $\mu \in [\mu_1,\mu_2]$
where $l_1,\mu_1>0$, $\delta_1=l_2+1-l_1$, $\delta_2=\mu_2+1-\mu_1$, $\delta_1\ll l_1$, $\delta_2\ll \mu_1$
$\mu_2<l_1$ and $\mu_1\gg (l_1-\mu_1)$. This is the state where the quantity $\mu$ is close to its maximum value $l$.
As follows from Eqs. (\ref{orbmagn}) and (\ref{L}), in this state the quantity $L_z$ is much 
greater than $L_x$ and $L_y$ and, as 
follows from Eq. (\ref{FGsemicl}), the quantities $F_z$ and $G_z$ are small. So on classical level this state
describes a motion of the particle in the $xy$ plane. The quantity $L_z$ in this state is obviously semiclassical 
since $\chi(l,\mu)$ is the eigenvector of the operator $L_z$ with the eigenvalue $\mu$. As follows from 
Eq. (\ref{FGsemicl}), the action of the operators $(F_+,F_-,G_+,G_-)$ on this state can be described by the
following approximate formulas:
\begin{eqnarray}
&&F_+\chi(l,\mu)=-\frac{il_0}{2}\chi(l-1,\mu-1),\quad F_-\chi(l,\mu)=\frac{il_0}{2}\chi(l+1,\mu+1)\nonumber\\
&&G_+\chi(l,\mu)=\frac{1}{2}\chi(l-1,\mu-1),\quad G_-\chi(l,\mu)=\frac{1}{2}\chi(l+1,\mu+1)
\label{FGsimple}
\end{eqnarray}
where $l_0$ is a value from the interval $[l_1,l_2]$.

Consider a simple model when $\chi(l,\mu)=exp[i(l\alpha -\mu\beta)]/(\delta_1\delta_2)^{1/2}$, $l\in [l_1,l_2]$
and $\mu\in [\mu_1,\mu_2]$. Then a simple direct calculation using Eq. (\ref{FGsimple}) gives
\begin{eqnarray}
&&{\bar G}_x=cos\gamma,\quad {\bar G}_y=-sin\gamma\quad {\bar F}_x=-l_0sin\gamma\quad {\bar F}_y=-l_0cos\gamma\nonumber\\
&& \Delta G_x=\Delta G_y=(\frac{1}{\delta_1}+\frac{1}{\delta_2})^{1/2},\quad 
\Delta F_x=\Delta F_y=l_0(\frac{1}{\delta_1}+\frac{1}{\delta_2})^{1/2}
\label{meanGF}
\end{eqnarray}
where $\gamma=\alpha-\beta$. Hence the vector quantities ${\bf F}$ and ${\bf G}$ are semiclassical since either 
$|cos\gamma|$ or $|sin\gamma|$ or both are much greater than $(\delta_1+\delta_2)/(\delta_1\delta_2)$.

\section{New position operator and wave packet spreading}
\label{newWPS}

If the space of states is implemented according to the scalar product (\ref{newscalar}) then the dependence 
of the wave function on $t$ is
\begin{equation}
\chi(p, k,\mu,t)=exp[-\frac{i}{\hbar}(m^2c^2+p^2)^{1/2}ct]\chi(p, k,\mu,t=0)
\label{chitnew}
\end{equation}
As noted in Secs. \ref{NRWPS} and \ref{RelWPS}, there is no WPS in momentum space and this is natural in view of momentum conservation.
Then, as already noted, the distribution of the quantity ${\bf r}_{\bot}$ does not depend on time and
this is natural from the considerations described in Sec. \ref{consistent}.

At the same time, the dependence of the $r_{||}$ distribution on time can be calculated in full analogy with
Sec. \ref{NRWPS}. Indeed, consider, for example a function $\chi(p,l, \mu, t=0)$ having the form
\begin{equation}
\chi(p,l, \mu, t=0)=\chi(p,t=0)\chi(l,\mu)
\label{factoriz}
\end{equation}
Then, as follows from Eqs. (\ref{psir||in}) and (\ref{chitnew}),
\begin{equation}
\psi(r,t)=\int exp[-\frac{i}{\hbar}(m^2c^2+p^2)^{1/2}ct+\frac{i}{\hbar}pr]\chi(p, t=0)\frac{dp}{(2\pi\hbar)^{1/2}}
\label{psir||int}
\end{equation}

Suppose that the function $\chi(p, t=0)$  is given by Eq. (\ref{pr||}). Then in full analogy with the calculations
in Sec. \ref{NRWPS} we get that in the nonrelativistic case the $r_{||}$ distribution is defined by the wave
function
\begin{equation}
\psi(r,t)=\frac{1}{\pi^{1/4}b^{1/2}}(1+\frac{i\hbar t}{mb^2})^{-1/2}exp[-\frac{(r-r_0-v_0t)^2}{2b^2(1+\frac{\hbar^2t^2}{m^2b^4})}(1-\frac{i\hbar t}{mb^2})+
\frac{i}{\hbar}p_0r-\frac{ip_0^2t}{2m\hbar}]
\label{psirtnew}
\end{equation}
where $v_0=p_0/m$ is the classical speed of the particle in the direction of the particle momentum. 
Hence the WPS effect in this direction is similar to that given by Eq. (\ref{psirt}) in standard theory.

In the opposite case when the particle is ultrarelativistic, Eq. (\ref{psir||int}) can be written as
\begin{equation}
\psi(r,t)=\int exp[\frac{i}{\hbar}p(r-ct)]\chi(p, t=0)\frac{dp}{(2\pi\hbar)^{1/2}}
\label{ultra}
\end{equation}
Hence, as follows from Eq. (\ref{psir||B}):
\begin{equation}
\psi(r,t)=\frac{1}{\pi^{1/4}b^{1/2}}exp[-\frac{(r-r_0-ct)^2}{2b^2}+\frac{i}{\hbar}p_0(r-ct)]
\label{psir||C}
\end{equation}
In particular, for an ultrarelativistic particle there is no WPS in the direction of particle momentum and
this is in agreement with the results of Sec. \ref{RelWPS}.

We conclude that in our approach an ultrarelativistic particle (e.g. the photon) experiences WPS neither in the
direction of its momentum nor in perpendicular directions, i.e. the WPS effect for an ultrarelativistic particle
is absent at all.

Let us note that the absence of WPS in perpendicular directions is simply a consequence of the fact that 
a consistently defined operator ${\bf {\cal R}}_{\bot}$ commutes with the Hamiltonian.
In quantum theory a physical quantity is called conserved if its operator commutes with the Hamiltonian.
Therefore ${\bf r}_{\bot}$ is a conserved physical quantity. In contrast to classical theory, this 
does not mean that ${\bf r}_{\bot}$ should necessarily have only one value but means
that the ${\bf r}_{\bot}$ distribution does not depend on time.  
 On the other hand, the longitudinal coordinate is not a conserved physical quantity
since a particle
is moving along the direction of its momentum. However, in a special case of ultrarelativistic particle
the absence of WPS is simply a consequence of the fact that the wave function given by Eq. (\ref{ultra})
depends on $r$ and $t$ only via a combination of $r-ct$.

\section{Discussion and conclusion}
\label{conclusion}

In the present paper we consider a problem of constructing position operator in quantum theory. As noted in Sec.
\ref{intropos}, this operator is needed in situations where semiclassical approximation works with a high
accuracy. 

A standard choice of the position operator in momentum space is 
$i\hbar\partial/\partial{\bf p}$. A motivation for this choice is discussed in Sec. \ref{classical}.
We note that this choice is not consistent since $i\hbar\partial/\partial p_j$
cannot be a physical position operator in directions where the momentum is small. 
Physicists did not pay attention
to the inconsistency probably for the following reason: as explained in textbooks, 
transition from quantum to classical theory can be performed such that if the 
coordinate wave
function contains a rapidly oscillating exponent $exp(iS/\hbar)$ where $S$ is the classical action then in the
formal limit $\hbar\to 0$ the Schr\"{o}dinger equation becomes the Hamilton-Jacobi equation. 

However, an inevitable
consequence of standard quantum theory is the effect of wave packet spreading (WPS). This fact has not been
considered as a drawback of the theory. Probably the reasons are that for macroscopic bodies this effect 
is extremely small while in
experiments on the Earth with atoms and elementary particles spreading probably does not have
enough time to manifest itself. However, for photons traveling to the Earth from distant objects this effect 
is considerable, and it seems that this fact has been overlooked by physicists.

As shown in Sec. \ref{experiment}, if the WPS effect for photons traveling to Earth from distant objects
is as given by standard theory then we have several fundamental paradoxes. The most striking of
them is that standard theory contradicts our experience on observations of stars. 

We propose a new definition of the position operator which we treat as consistent for the following reasons.
Our position operator is defined by two components - in the direction along the momentum
and in perpendicular directions. The first part has a familiar form $i\hbar \partial/\partial p$ and is
treated as the operator of the longitudinal coordinate if the magnitude of
$p$ is rather large. At the same condition the position operator in the perpendicular directions
is defined as a quantum generalization of the relation ${\bf r}_{\bot}\times {\bf p}={\bf L}$.
So in contrast to the standard definition of the position operator, the new operator is expected
to be physical only if the {\it magnitude} of the momentum is rather large. 

As a consequence of our construction, WPS in directions perpendicular to the particle
momentum is absent regardless of whether the particle is nonrelativistic or relativistic. 
Moreover, for an ultrarelativistic particle the effect of WPS is absent at all. 

Different components of the new position operator commute with each other only in the formal limit $\hbar\to 0$. 
As a consequence, there is no wave function in coordinate representation. In particular, there is no
quantum analog of the coordinate Coulomb potential (see the discussion in Sec. \ref{intropos}). A possibility 
that coordinates can be noncommutative has been first
discussed by Snyder \cite{Snyder} and it is implemented in several modern theories. In those theories the
measure of noncommutativity is defined by a parameter $l$ called the fundamental length  (the role of which can be
played e.g. by the Planck length or the Schwarzschild radius). In the formal limit $l\to 0$ the coordinates become 
standard ones related to momenta by a Fourier transform. 
As shown in the present paper, this is unacceptable in view of the WPS paradoxes. One of ideas of those theories
is that with a nonzero $l$ it might be possible to resolve difficulties of standard theory where $l=0$ (see
e.g. Ref. \cite{Smolin} and references therein). At the same time, in our approach 
there can be no notion of fundamental length since commutativity of coordinates takes place only in the
formal limit $\hbar\to 0$.

The absence of the coordinate wave function is not unusual. For example, there is no wave function
in the angular momentum representation because different components 
of the angular momentum operator commute only in the formal limit $\hbar\to 0$. 
However, on classical level all the commutators can be neglected and different components
of the position vector and angular momentum can be treated independently.

In our approach the uncertainties of each component of the photon momentum and each component of the photon coordinate do not change with time. If in some problem those quantities can be treated as small then the photon can be treated as a pointlike particle moving along classical trajectory. 
So in our approach the coordinate photon wave function never has a cosmic size and there
can be no paradoxes discussed in Sec. \ref{experiment}.

In view of the absence of the coordinate wave function, such quantum problems as diffraction and interference of
single photon should be considered only in momentum representation. In particular, if boundary conditions are needed
they should be formulated in that representation. When a problem is solved and characteristic spatial dimensions in
the problem are greater than uncertainties of all the coordinates one can discuss spatial features of the process.
   
As noted in Sec. \ref{WPW}, in  standard quantum theory photons
comprising a classical electromagnetic wave packet cannot be (approximately) treated as pointlike particles 
in view of the WPW paradox. However, in our approach, in view of the absence of WPS for massless particles, 
the usual intuition is restored
and photons comprising a divergent classical wave packet can be (approximately) treated as pointlike particles.
Moreover, the phenomenon of divergence of a classical wave packet can now be naturally explained simply as
a consequence of the fact that different photons in the packet have different momenta.

Our consideration also poses
a problem whether the results of classical electrodynamics can be applied for wave packets moving for a long period
of time. For example, as noted in Sec. \ref{experiment}, even classical theory predicts that when a
wave packet emitted in a gamma-ray burst or by a pulsar reaches the Earth, the width of the packet 
is extremely large (while the value predicted
by standard quantum theory is even much greater) and this poses a problem whether such a packet can be detected. 
A natural explanation of why classical theory does not apply in this case follows. 
As noted in Sec. \ref{momentum}, classical electromagnetic fields should be understood as a result of taking mean
characteristics for many photons. Then the fields will be (approximately) continuous if the density of the photons
is high. However, for a divergent beam of photons their density decreases with time. Hence after a long period of
time the mean characteristics of the photons in the beam cannot represent continuous fields. In other words, in this
situation the set of photons cannot be effectively described by classical electromagnetic fields.

The new position operator might also have applications in the problem of neutrino oscillations. As pointed out by several authors 
(see e.g. Refs. \cite{Beuthe,Akhmedov,Naumov}) this problem should be considered from the point of
view that for describing observable neutrinos one should treat them as quantum superpositions of wave packets with
different neutrino flavors. Then the choice of the position operator might play an important role.
 
The position operator proposed in the present paper is also important in view of the following. There exists a wide 
literature discussing the Einstein-Podolsky-Rosen paradox, locality in quantum theory, quantum entanglement, Bell's 
theorem and similar problems (see e.g. Ref. \cite{Griffiths} and references therein). Consider, for example, the
following problem in standard theory. Let at $t=0$ particles 1 and 2 be localized inside finite volumes $V_1$ and $V_2$,
respectively, such that the volumes are very far from each other. Hence the particles don't interact with each other.
However, as follows from Eq. (\ref{trel}), their wave functions will overlap at any $t>0$ and hence the interaction 
can be transmitted even with an infinite speed. This is often characterized as quantum nonlocality, entanglement and/or
action at a distance. 

Consider now this problem in the framework of our approach. Since in this approach there is no wave function in
coordinate representation, there is no notion of a particle localized inside a finite volume.
Hence a problem arises whether on quantum level the notions of locality or nonlocality have a physical meaning. 
In addition, spreading does not take place in directions perpendicular to the particle momenta and for ultrarelativistic
particles spreading does not occur at all. Hence, at least in the case of ultrarelativistic particles, this kind of 
interaction does not occur in agreement with classical intuition that no interaction can be
transmitted with the speed greater than $c$. This example poses a problem whether the position operator should be
modified not only in directions perpendicular to particle momenta but also in longitudinal directions such that
the effect of WPS should be excluded at all.

A problem discussed in a wide literature is whether evolution of a quantum system
can be always described by the time dependent Schr\"{o}dinger equation. We will discuss this problem in view of the
statements (see e.g. Refs. \cite{Rovelli,Keaton}) that $t$ cannot be treated as a fundamental
physical quantity. The reason is that all fundamental physical
laws do not require time and the quantity $t$ is obsolete on fundamental level. A
hypothesis that time is an independently flowing fundamental continuous quantity has
been first proposed by Newton. However, a problem arises whether this hypothesis is
compatible with the principle that the definition of a physical quantity is a description
of how this quantity can be measured.

Consider first the problem of time in classical mechanics. A standard
treatment of this theory is that its goal is to solve equations of motion and get
classical trajectories where coordinates and momenta are functions of $t$. In 
Hamiltonian mechanics the action can be written as $S = S_0-\int Hdt$ where $S_0$ does not 
depend on $t$ and is called the abbreviated action. Then, as
explained in textbooks, the dependence of the coordinates and momenta on
$t$ can be obtained from a variational principle with the action $S$. Suppose now that
one wishes to consider a problem which is usually treated as less general: to find
not the dependence of the coordinates and momenta on $t$ but only possible forms of
trajectories in the phase space without mentioning time at all. If the energy is
a conserved physical quantity then, as described in textbooks, this problem
can be solved by using the Maupertuis principle involving only $S_0$.

However, the latter problem {\it is not} less general than the former one. For
illustration we first consider the one-body case. Here the phase space can be described
by the quantities $(r_{||},{\bf r}_{\bot},{\bf G},p)$ discussed in Sec. \ref{consistent}. 
Suppose that by using the Maupertuis
principle one has solved the problem with some initial values of coordinates
and momenta. One can choose $r_{||}$ such that it is zero at the initial point and increases
along the trajectory. Then $r_{||} = s$ where $s$ is the length along the spacial trajectory
and a natural parametrization for the trajectory in the phase space is such that
$({\bf r}_{\bot},{\bf G},p)$ are functions of $r_{||}=s$. This is an additional indication that 
our choice of the position operator is more natural than 
standard one. At this stage the problem does not contain $t$ yet. We
can note that in standard case $ds/dt=|{\bf v}(s)|=|{\bf p}(s)|/E(s)$. Hence
in the problem under consideration one can {\it define} $t$ such that $dt = E(s)ds/|{\bf p}(s)|$ and
hence the value of $t$ at any point of the trajectory can be obtained by integration. In
the case of many bodies one can define $t$ by using the spatial trajectory of any body
and the result does not depend on the choice of the body. Hence 
the general problem of classical mechanics can be formulated without mentioning $t$ while if one wishes to
work with $t$ then, by definition, this value can flow only in positive direction.

Consider now the problem of time in quantum theory. In the case of
one strongly quantum system (i.e. the system which cannot be described in classical
theory) a problem arises whether there exists a quantum analog of the Maupertuis
principle and whether time can be defined by using this analog. This is a difficult
unsolved problem. A possible approach for solving this problem has been proposed
in Ref. \cite{Rovelli}. However, one can consider a situation when a quantum system under
consideration is a small subsystem of a big system where the other subsystem - the environment, is
strongly classical. Then one can define $t$ for the environment as described
above. The author of Ref. \cite{Keaton} considers a scenario when the system as a whole is
described by the stationary Schr\"{o}dinger equation $H\Psi	 = E\Psi$	 but the small quantum
subsystem is described by the time dependent Schr\"{o}dinger equation where $t$ is
defined for the environment as $t=\partial S_0/\partial E$.

One might think that this scenario gives a natural solution of the problem of time in quantum theory.
Indeed, in this scenario it is clear why a quantum system is described by the  
Schr\"{o}dinger equation depending on the classical parameter $t$ which is not an operator: because 
$t$ is the physical quantity
characterizing not the quantum system but the environment. This scenario seems also natural because
it is in the spirit of the Copenhagen interpretation of quantum theory: the evolution of a quantum system
can be characterized only in terms of measurements which in the Copenhagen interpretation are treated as
interactions with classical objects. However, this scenario encounters the following problems. As noted
in Ref. \cite{Keaton}, it does not solve the problem of quantum jumps. For example, as noted in Sec.
\ref{momentum}, the 21cm transition in the hydrogen atom cannot be described by the evolution operator
depending on the continuous parameter $t$. Another problem is that the environment can be a classical
object only in some approximation and hence $t$ can be only an approximately continuous parameter. 
Finally, the Copenhagen interpretation cannot be universal in all situations. For example, if the Big
Bang hypothesis is correct then at the early stage of the Universe there were no classical objects but
nevertheless physics should somehow describe evolution even in this situation. 

Our result for ultrarelativistic particles can be treated as ideal: quantum
theory reproduces the motion along a classical trajectory without any spreading. However, this is only a
special case of one free elementary particle. If quantum theory is treated as more general than the classical one
then it should describe not only elementary particles and atoms but even the motion of macroscopic bodies in the Solar
System and in the Universe. We believe that the 
assumption that the evolution of macroscopic bodies can be described by the Schr\"{o}dinger
equation is unphysical. For example, if the motion of the Earth is described by the 
evolution operator $exp[-iH(t_2-t_1)/\hbar]$ where $H$ is the Hamiltonian of the Earth then the quantity 
$H(t_2-t_1)/\hbar$ becomes of the order of
unity when $t_2-t_1$ is a quantity of the order of $10^{-68}s$ if the Hamiltonian is written in 
nonrelativistic form and $10^{-76}s$ if it is written in relativistic form. 
Such time intervals seem to be unphysical and so in the given case
the approximation when $t$ is a continuous parameter seems to be unphysical too. In modern 
theories (e.g. in the Big Bang hypothesis) it is often stated that the Planck time $t_P\approx 10^{-43}s$ 
is a physical minimum time interval. However, at present
there are no experiments confirming that time intervals of the order of $10^{-43}s$ can be measured.

The time dependent Schr\"{o}dinger equation has not been experimentally verified and the major theoretical
arguments in favor of this equation are as follows: a) the Hamiltonian is the generator of the time translation
in the Minkowski space; b) this equation becomes the Hamilton-Jacobi one in the formal limit $\hbar\to 0$.
However, as noted in Sec. \ref{intropos}, quantum theory should not be based on the space-time background
and the conclusion b) is made without taking into account the WPS effect. Hence the problem of describing evolution in quantum theory remains open.

Let us now return to the problem of the position operator.
As noted above, in directions perpendicular to the particle momentum the choice of the position
operator is based only on the requirement that semiclassical approximation should reproduce the standard 
relation ${\bf r}_{\bot}\times{\bf p}={\bf L}$. This requirement 
seems to be beyond any doubts since {\it on classical level} this relation is confirmed in numerous experiments. 
At the same time, the choice $i\hbar \partial/\partial p$
of the coordinate operator in the longitudinal direction is analogous to that in standard theory and hence
one might expect that this operator is physical if the magnitude of $p$ is rather large (see, however, the above
remark about the entanglement caused by WPS).

It will be shown in a separate publication that the construction of the position operator described in this
paper for the case of Poincare invariant theory can be generalized to the case of de Sitter (dS) invariant theory.
In this case the interpretation of the position operator is even
more important than in Poincare invariant theory. The reason is that even the free two-body mass operator
in the dS theory depends not only on the relative two-body momentum but also on the distance between the
particles. 

As argued in Ref. \cite{gravity}, in dS theory over a Galois field the assumption that the dS analog of the operator
$i\hbar \partial/\partial p$ is the operator of the longitudinal coordinate is not valid 
{\it for macroscopic bodies} 
(even if $p$ is large) since in that case semiclassical approximation is not valid. We have proposed a 
modification of the
position operator such that quantum theory reproduces for the two-body mass operator the mean value  
compatible with the Newton law of gravity. Then a problem arises
how quantum theory can reproduce classical evolution for macroscopic bodies. 

The above examples show that at macroscopic level a consistent definition of the transition from quantum to 
classical theory is the fundamental open problem.

\begin{center} {\bf Acknowledgements} \end{center}

I am very grateful to Anatoly Kamchatnov for pointing out that the conclusion about the momentum distribution of the photon emitted by a star and detected on the Earth was erroneous. As a consequence, statements about some paradoxes were erroneous too. His critics of my approach as a whole was also very stimulating. I am also grateful to Steven Carlip, Philip Gibbs, Mikhail Ivanov, Gregory Keaton, Volodya Netchitailo, Carlo Rovelli and the anonymous referee  for important remarks.

\end{document}